%% file: main.tex
\documentclass[sigconf]{acmart}
\AtBeginDocument{%
  \providecommand\BibTeX{{%
    \normalfont B\kern-0.5em{\scshape i\kern-0.25em b}\kern-0.8em\TeX}}}

\copyrightyear{2021} 
\acmYear{2021} 
\setcopyright{rightsretained} 
\acmConference[SC '21]{The International Conference for High Performance Computing, Networking, Storage and Analysis}{November 14--19, 2021}{St. Louis, MO, USA}
\acmBooktitle{The International Conference for High Performance Computing, Networking, Storage and Analysis (SC '21), November 14--19, 2021, St. Louis, MO, USA}
\acmDOI{10.1145/3458817.3476157}
\acmISBN{978-1-4503-8442-1/21/11}

\newcommand*\Mname{\textsc{APNN-TC}}

\usepackage[ruled,vlined,linesnumbered]{algorithm2e}

\SetCommentSty{mycommfont}

\usepackage{xcolor}
\usepackage{color}
\definecolor{codegreen}{rgb}{0,0.6,0}
\definecolor{codegray}{rgb}{0.5,0.5,0.5}
\definecolor{codepurple}{rgb}{0.58,0,0.82}
\definecolor{backcolour}{rgb}{0.95,0.95,0.92}
\definecolor{textblue}{rgb}{.2,.2,.7}
\definecolor{textred}{rgb}{0.54,0,0}
\definecolor{textgreen}{rgb}{0,0.43,0}
\definecolor{codered}{rgb}{201,72,12}

\usepackage[T1]{fontenc}
\usepackage[scaled=0.85]{beramono} 
\usepackage{listings}
\usepackage{multirow}

\lstdefinestyle{tt1}{
language=C,
basicstyle=\linespread{0.9}\ttfamily\footnotesize,
breaklines=true,
numbers=left,
frame=single,
numberstyle=\tiny, 
stepnumber=1,
numbersep=5pt, 
tabsize=4,
keywordstyle=\bfseries\color{codegreen},
commentstyle=\color{textred},   
stringstyle=\color{textgreen},
columns=fullflexible,
keepspaces=true,
xleftmargin=\parindent,
showstringspaces=false,
otherkeywords = {True, False},
keywordstyle=[2]\color{codepurple}\bfseries,
keywords=[2]{wmma, b1},
keywordstyle=[3]\color{textblue}\bfseries,
keywords=[3]{fragment, load_matrix_sync, mma_sync, store_matrix_sync},
keywordstyle=[4]\color{codegreen}\bfseries,
keywords=[4]{matrix_a, row_major, column_major, mem_row_major, tf32},
}
\usepackage{booktabs}
\usepackage{subcaption}
\usepackage{graphicx}
\usepackage{amsmath}
\usepackage{multirow}
\usepackage{algpseudocode}
\usepackage{subcaption}

\usepackage{mathtools}

\usepackage[title]{appendix}

\usepackage{lipsum}

\newcommand\blfootnote[1]{%
  \begingroup
  \renewcommand\thefootnote{}\footnote{#1}%
  \addtocounter{footnote}{-1}%
  \endgroup
}
\usepackage{fancyhdr}
\usepackage{fvextra}
\usepackage[utf8]{inputenc}

\begin{document}

\title{APNN-TC: Accelerating Arbitrary Precision \\ Neural Networks on Ampere GPU Tensor Cores}


\author[B. Feng et al.]{Boyuan Feng\texorpdfstring{$^\dagger$}{dagger}$^\diamondsuit$, Yuke Wang\texorpdfstring{$^\dagger$}{dagger}$^\diamondsuit$, Tong Geng*, Ang Li*, Yufei Ding\texorpdfstring{$^\dagger$}{dagger}}
\affiliation{
  \institution{\texorpdfstring{$^\dagger$}{dagger}\{boyuan, yuke\_wang, yufeiding\}@cs.ucsb.edu, *\{tong.geng, ang.li\}@pnnl.gov}
  \institution{\texorpdfstring{$^\dagger$}{dagger}University of California, Santa Barbara, \\ *Pacific Northwest National Laboratory.}
  \country{}
  }


\input{00_abstract}

\maketitle

\blfootnote{$^\diamondsuit$ The first two authors contribute equally.}

\input{01_intro}
\input{02_background}

\input{03_APComputation}

\input{04_APKernel}

\input{05_APNN}

\input{06_evaluation}
\input{07_Discussion}

\input{08_conclusion}

\vspace{-3pt}
\section{Acknowledgements}
We thank all anonymous reviewers for their valuable comments. This work was supported in part by NSF 1925717 and 2124039. This work was supported in part by the U.S. DOE Office of Science, Office of Advanced Scientific Computing Research, under award 66150: "CENATE - Center for Advanced Architecture Evaluation" and PNNL's Data-Model-Convergence (DMC) LDRD Initiative Computation-Flow-Architecture (CFA) project. The Pacific Northwest
National Laboratory is operated by Battelle for the U.S. Department of Energy under contract DE-AC05-76RL01830.

\bibliographystyle{ACM-Reference-Format}
\bibliography{sample-base}
\end{document}

%% file: 00_abstract.tex
\begin{abstract}
Over the years, accelerating neural networks with quantization has been widely studied.
Unfortunately, prior efforts with diverse precisions (e.g., 1-bit weights and 2-bit activations) are usually restricted by limited precision support on GPUs (e.g., int1 and int4).
To break such restrictions, we introduce the first Arbitrary Precision Neural Network framework (APNN-TC)\footnote{The project is open-sourced at https://github.com/BoyuanFeng/APNN-TC} to fully exploit quantization benefits on Ampere GPU Tensor Cores.
Specifically, APNN-TC first incorporates a novel emulation algorithm to support arbitrary short bit-width computation with int1 compute primitives and XOR/AND Boolean operations.
Second, APNN-TC integrates arbitrary precision layer designs to efficiently map our emulation algorithm to Tensor Cores with novel batching strategies and specialized memory organization.
Third, APNN-TC embodies a novel arbitrary precision NN design to minimize memory access across layers and further improve performance.
Extensive evaluations show that APNN-TC can achieve significant speedup over CUTLASS kernels and various NN models, such as ResNet and VGG.
\end{abstract}

%% file: 01_intro.tex
\vspace{-20pt}
\section{Introduction}
Over the recent years, demands to improve the performance of deep neural networks (DNNs) have never been satisfied. 
Prior work approaches faster and more efficient DNNs from different aspects, such as model pruning~\cite{liu2020autocompress, niu2020towards, ma2020pconv, dualInfer}, kernel factorization~\cite{mobilenet_2017_howard, sandler2018mobilenetv2, xception, DSXplore}, and data quantization~\cite{training-qnn, qcnn}.
Among those efforts, quantization-based DNN acceleration~\cite{searchlowbit,training-qnn,qcnn} finds its strengths in minimum modification of the original model architecture, lower memory consumption, and better runtime performance.

To accelerate quantized DNNs, many specialized cores have been introduced to support low-precision dense matrix-matrix multiplications, such as Tensor Processing Units (TPUs) \cite{jouppi2017datacenter}, Neural Network Processors (NNPs) \cite{hickmann2020intel}, and GPU Tensor Cores \cite{choquette2021nvidia}.
For example, NVIDIA introduces Tensor Cores in Volta architecture \cite{choquette2018volta} that support \texttt{FP16} matrix-matrix multiplication.
In Turing architecture, NVIDIA extends architecture support for more precisions (\textit{e.g.}, \texttt{int1} and \texttt{int4}) and bit-level operations (\textit{e.g.}, \texttt{XOR}) \cite{TPDS}.
Recently in the Ampere architecture, we find there is additional support for more precision and bit-level operations (\textit{e.g.}, \texttt{AND}).
However, these specialized cores still support a limited range of precisions with only architecture-level efforts, while quantized DNNs usually require arbitrary precisions (\textit{e.g.}, 1-bit weight and 2-bit activations).
In this paper, our key question is \textit{whether we can support arbitrary precision neural networks with the limited precisions on Tensor Cores}.



We identify two major challenges in accelerating arbitrary precision DNNs on Ampere GPU Tensor Cores.

\textbf{Lack of mathematical emulation design.}
To support arbitrary precisions (\textit{e.g.}, \texttt{int1} weights and \texttt{int2} activations), one naive approach is to represent these low-precision values with the supported high-precision values (\textit{e.g.}, \texttt{int4}).
However, this approach introduces extra overhead and prevents efficient quantized DNNs on Tensor Cores.
Another approach is to emulate with \texttt{int1} compute primitives.
However, with \texttt{int1} precision, Tensor Cores only support two bit-level operations (\textit{i.e.}, \texttt{XOR} and \texttt{AND}) and mathematical emulation designs are required to support multiplication and addition in quantized DNNs.
Moreover, quantized DNNs may have diverse input data (\textit{e.g.}, -1/+1 or 0/1), where different data may require different emulation designs.

\textbf{Lack of efficient implementation for arbitrary precision NN layers.}
To accelerate APNN on Tensor Cores, we need to efficiently map arbitrary precision NN layers to Tensor Cores with specialized compute primitives and memory architectures.
Existing works on accelerating binary neural networks simply split NN layers into small matrix tiles (\textit{e.g.}, $8\times 8$) to match Tensor Core compute primitives and improve the parallelism.
However, naively borrowing these strategies fails to exploit the data locality during NN layer computation especially for our emulation workload.
Moreover, arbitrary precision computation usually computes at the bit-level (\textit{e.g.}, \texttt{int3} or \texttt{int5}) while existing hardware devices such as CPUs and GPUs usually operate at the word or byte level.
Specialized bit operations and data organization are required to support efficient bit-level computation and avoid uncoalesced memory access.


\textbf{Lack of efficient NN framework designs.}
One standard approach to build quantized neural networks is to stack a sequence of NN layers, such as a convolution layer followed by a pooling layer and a quantization layer.
However, this approach ignores the data reuse opportunity across NN layers and leads to unnecessary memory overhead.
For example, on NNs with $n$ 2-bit activations, there are two semantic equivalent implementations -- quantization after reading 32-bit activations from the previous layer or quantization to 2-bit ones before writing to global memory for the next layer.
While these two implementations provide the same semantic, the former requires memory access of $32n$ bits while the latter only requires memory access of $2n$ bits.
\begin{figure} [t] \small
    \centering
    \includegraphics[width=0.97\linewidth]{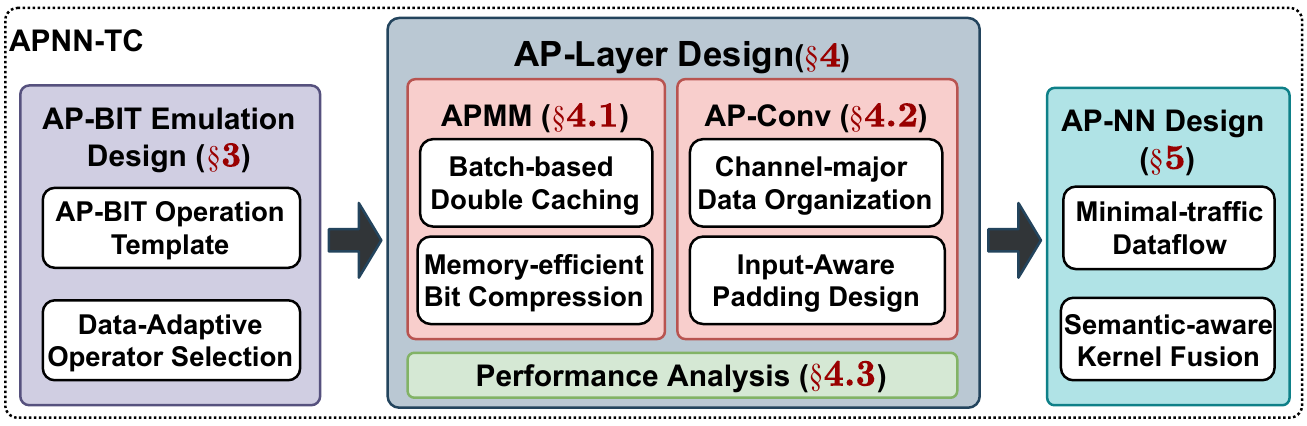}
    \vspace{-7pt}
    \caption{The overview of APNN framework.} 
    \label{fig: The overview of APNN framework.}
    \vspace{-15pt}
\end{figure}



To this end, we propose {\Mname} to accelerate Arbitrary Precision Neural Networks on Ampere GPU Tensor Cores, as illustrated in Figure \ref{fig: The overview of APNN framework.}.
First, we propose an \textit{AP-BIT emulation design} to support arbitrary-precision computation with 1-bit compute primitives.
Our AP-BIT algorithm can adaptively select operators (\textit{e.g.}, \texttt{XOR} or \texttt{AND}) to support diverse input data (\textit{e.g.}, -1/1 or 0/1).
Second, we build efficient \textit{AP-Layer design} including an arbitrary-precision matrix-matrix multiplication (APMM) layer for fully connected layers and an arbitrary-precision convolution (APConv) layer for convolution layers.
We propose a set of memory and computation designs (\textit{e.g.}, batch-based double caching and channel-major data organization) to fully exploit Tensor Core computation and minimize memory access.
We also incorporate a performance analysis to automatically tune the hyper-parameters in APMM and APConv.
Third, we propose an efficient \textit{APNN design} to improve the performance at the framework level.
It includes a minimal-traffic dataflow to support various precisions over APNN layers and a semantic-aware kernel fusion to minimize the data movement across layers.

In summary, we make the following contributions in this paper.
\begin{itemize}
    \item We develop APNN-TC to accelerate neural network on Ampere GPU Tensor Cores with arbitrary precision.
    \item We propose three novel techniques: a) an AP-BIT emulation design to support arbitrary-precision computation; b) an efficient AP-Layer design to achieve high performance at the layer level; c) an efficient APNN design to minimize the data movement across layers.
    \item Extensive experiments show that \Mname~can achieve up to $3.78\times$ speedup over CUTLASS kernels and $3.08\times$ speedup over CUBLAS kernels.
    \Mname~can also consistently outperform NNs implemented with built-in int8, half, or single precision. For example, with $2$-bit weights and $8$-bit activations, \Mname~can achieve more than $4\times$ latency reduction and $3\times$ higher throughput than the single-precision NN with only $2\%$ accuracy drop.

\end{itemize}

%% file: 02_background.tex
\vspace{-7pt}
\section{Related Works}
\subsection{APNN algorithm designs}
Arbitrary precision (lower than INT8) neural network (APNN) algorithms have been widely studied \cite{HanMD15,TPDS,BinaryConnect,DoReFa,zhang2018lq,HAQ,OLCEL,li2019bstc, geng2020o3bnn} to fully explore the spectrum of NN performance and NN accuracy and cater to diverse application requirements.
In addition to widely supported precisions on modern GPUs (\textit{e.g.}, \texttt{int1}, \texttt{int4}, and \texttt{int8}), these APNNs usually utilize more diverse precisions such as \texttt{int2}, \texttt{int3}, and \texttt{int5}.
APNNs may also have different precisions for weights and activations (\textit{e.g.}, 1-bit weights and 2-bit activations).
Comparing with INT8 quantized neural networks, APNNs provide better performance and memory efficiency at the cost of (slightly) degraded accuracy.
Popular APNNs include DoReFa-Net \cite{DoReFa} for 1-bit weights and 2-bit activations, LQ-Nets \cite{zhang2018lq} for 1-4 bits, HAQ \cite{HAQ} for 1-8 bits, OLAccel \cite{OLCEL} for 4 bits, BSTC \cite{li2019bstc} and TCBNN \cite{TPDS} for 1 bits.
In this paper, we follow LQ-Nets \cite{zhang2018lq} that starts from a full-precision NN and adopts the quantization error minimization (QEM) strategy to generate quantized NNs.

\vspace{-7pt}
\subsection{APNN Hardware Supports}
While many APNN algorithms have been designed, the hardware supports are still limited.
One direction is to build FPGA and ASIC based implementations \cite{HAQ, OLCEL} to demonstrate the performance benefits of APNNs.
However, these implementations usually require specialized hardware designs to support arbitrary-precision computation and cannot be applied to GPUs.
Another direction is to utilize built-in precisions on GPUs for quantized neural networks.
Taking the most famous Pytorch~\cite{pytorch} framework as an example, it supports FP32, FP16, and BF16 models on GPUs and int8 quantization on x86 CPUs with AVX2 support.
Recently, BSTC \cite{li2019bstc} and BTC \cite{TPDS} accelerates binary neural networks on GPUs by exploiting the int1 compute primitive.
However, existing works can only build on the limited precision supported on GPUs (\textit{e.g.}, \texttt{int1}, \texttt{int4}, and \texttt{int8}) and cannot fully exploit the performance benefits from APNNs.
In this paper, we build the first generalized framework to accelerate arbitrary-precision neural networks on Ampere GPU Tensor Cores.

\vspace{-7pt}
\subsection{Tensor Cores}
Tensor Cores are specialized cores for accelerating neural networks in terms of matrix-matrix multiplications.
Tensor Cores are introduced in recent NVIDIA GPUs since Volta architecture \cite{volta}.
Different from CUDA Cores that compute scalar values with individual threads, Tensor Cores compute at the matrix level with all threads in a warp \cite{raihan2019modeling}.
For example, the 1-bit Tensor Core compute primitive takes two \texttt{int1} input matrices A and B of shape $8\times128$ and generates an \texttt{int32} output matrix C of shape $8\times 8$ \cite{TPDS}.
In Volta architecture, Tensor Cores support only half-precision computation \cite{jia2018dissecting}.
To support more quantized neural networks, Tensor Cores add more precisions including \texttt{int1}, \texttt{int4}, and \texttt{int8} in Turing architecture \cite{jia2019dissecting}.
Regarding \texttt{int1} precision, Tensor Cores support only \texttt{XOR} logical operation in Turing architecture and recently add \texttt{AND} logical operation in Ampere architecture \cite{ampere}.
Despite these hardware efforts on supporting more precisions, arbitrary precisions are still not supported.
This is the first work to support arbitrary precision computation on Ampere GPU Tensor Cores with \texttt{int1} precision and support for both \texttt{XOR} and \texttt{AND} operations.

\begin{figure*}
    \centering
    \includegraphics[width=0.8\textwidth]{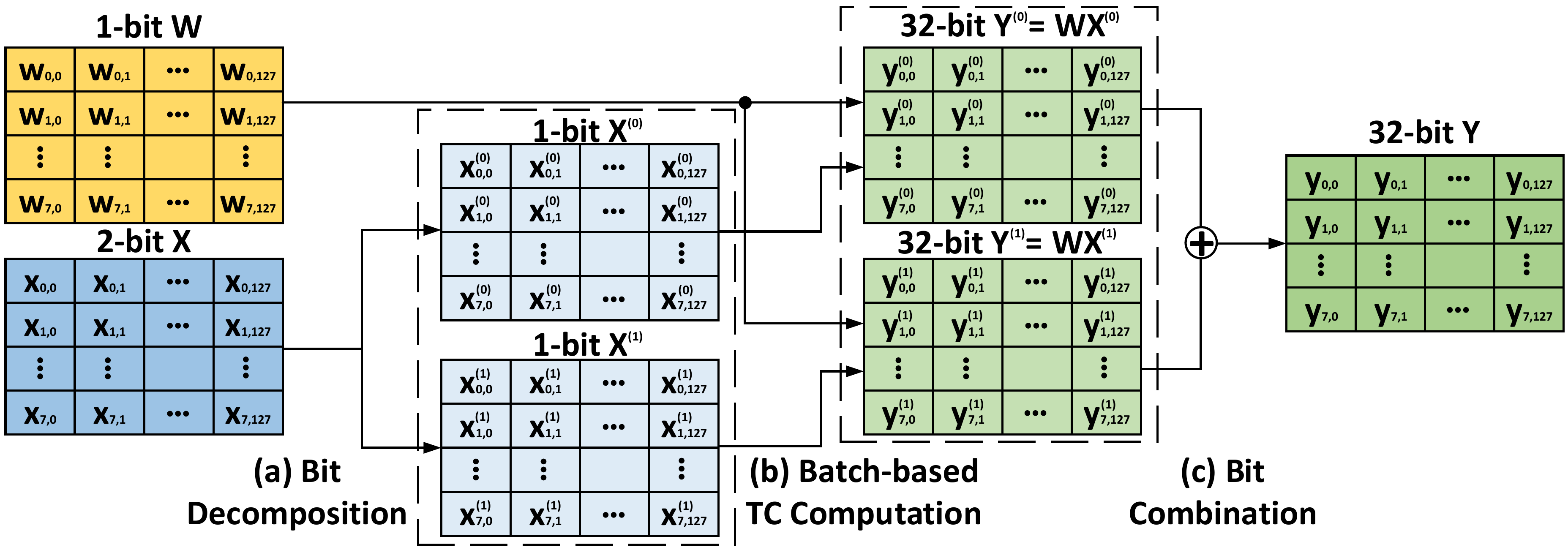}
    \vspace{-5pt}
    \caption{Illustration of AP-Bit Operation Template with 1-bit weight W and 2-bit feature X, which can be generalized to arbitrary weight bits and feature bits. Note that $X^{(0)}$ and $X^{(1)}$ in the dashed box are batched into a single large matrix during computation, which will be discussed in Section \ref{sec:APMM}.}
    \vspace{-10pt}
    \label{fig:BIT-Illustration}
\end{figure*}

%% file: 03_APComputation.tex

\section{AP-Bit Emulation Design}

In this section, we design an AP-BIT emulation on Tensor Cores to support arbitrary-precision computation.
We first design an AP-Bit operation template that supports arbitrary-precision computation with 1-bit compute primitive on Tensor Cores.
Then, we propose a data adaptive operator selection to automatically support various input data (\textit{e.g.}, -1/+1 and 0/1) with bitwise \texttt{XOR} and \texttt{AND} on Tensor Cores.
Here, we focus on the algorithm design on small matrices (\textit{i.e.}, input matrices of $8\times 128$ and output matrix of $8\times8$) that can fit directly on Tensor Core compute primitives.
We will discuss the efficient computation of large matrices in the next section.

\subsection{AP-Bit Operation Template Design}

The AP-Bit operation template takes a matrix $W$ with $p$-bit elements and a matrix $X$ with $q$-bit elements, and computes with 1-bit operations on Tensor Cores to generate a $32$-bit output matrix $Y = WX$.
Our key observation is that each arbitrary-bit scalar digit can be decomposed to a sequence of 1-bit scalar digits and the arbitrary computation can be conducted with only 1-bit operations and shift operations.
Formally, to support scalar-level arbitrary precision computation $wx$ of a 1-bit weight $w$ and a 2-bit feature $x = x^{(1)}x^{(0)}$ with $w, x^{(i)} \in $ \texttt{int1}, we can first decompose 1-bit values $x^{(1)}$ and $x^{(0)}$ from the 2-bit feature $x$ as:
\begin{equation*} \small
    x^{(1)} = (x \gg 1) \& 1, \quad x^{(0)} = (x \gg 0) \& 1
\end{equation*}
Suppose we have an 1-bit operation $OP(a,b)$ (\textit{e.g.}, the \texttt{bmma} API of Tensor Cores) that takes 1-bit inputs and generate 32-bit outputs, we can compute $wx$ as
\begin{equation*} \small
    wx = OP(w,x^{(1)})*2 + OP(w,x^{(0)})
\end{equation*}


We illustrate our AP-Bit operation template in Figure \ref{fig:BIT-Illustration}.
Here, we focus on a 1-bit weight matrix $W$ of shape $8\times 128$ and a 2-bit feature matrix $X$ of shape $8\times 128$ to illustrate our algorithm design.
A naive approach is to use 4-bit integers to represent each 1-bit element $w_{i,j}$ and 2-bit element $x_{i,j}$, and then use the $int4$ compute primitive on Tensor Cores.
However, this approach would lead to unnecessary memory and computation overhead.
Instead, we propose to exploit the $int1$ compute primitive on Tensor Cores to support arbitrary-precision computation by dynamically adjusting the memory and computation requirement.
In particular, the first step is to conduct \textbf{bit decomposition} by splitting a 2-bit $x_{i,j}$ to two 1-bit elements $x_{i,j}^{(0)}$ and $x_{i,j}^{(0)}$:
\begin{equation*} \small
    x_{i,j}^{(1)} = (x_{i,j}\gg1) \& 1, \quad x_{i,j}^{(0)} = (x_{i,j} \gg 0) \& 1
\end{equation*}
These 1-bit elements are then packed into 1-bit matrix $X^{0}$ and $X^{(1)}$.
The second step is to conduct \textbf{batch-based Tensor Core computation} on these 1-bit matrices with the \texttt{bmma} API and generate 32-bit output matrices
\begin{equation*} \small
    Y^{(0)} = \text{\texttt{bmma}}(W, X^{(0)}), \quad Y^{(1)} = \text{\texttt{bmma}}(W, X^{(1)})
\end{equation*}
These matrices can be computed directly with the \texttt{bmma} API since all of them have the shape of $8\times 128$.
We also note that Tensor Core primitives for \texttt{int1}, \texttt{int4}, and \texttt{int8} generate 32-bit output matrices to accumulate a large number of bit-operation outputs and avoid overflow.
The third step is to conduct \textbf{bit combination} and generate the final output matrix $Y$
\begin{equation} \label{eq:bitCombination} \small
    Y_{i,j} = Y_{i,j}^{(1)}*2 + Y_{i,j}^{(0)}
\end{equation}
Here, $Y_{i,j}$, $Y_{i,j}^{(1)}$ and $Y_{i,j}^{(0)}$ refer to the $(i,j)^{th}$ scalar elements of matrix $Y$, $Y^{(1)}$ and $Y^{(0)}$, respectively.
For notation simplicity, we abbreviate Equation \ref{eq:bitCombination} as $Y = Y^{(1)}*2 + Y^{(0)}$ in the following sections to represent the scalar multiplication and elementwise addition.
We note that $Y=WX$ mathematically.

It is not hard to see that this computation can be generalized to matrices with arbitrary bits $p$ and $q$.
Formally, given a $p$-bit weight matrix $W$ and a $q$-bit weight matrix $X$, we can first decompose into 1-bit matrices $W^{(s)}, s\in \{0,1,...,p-1\}$ and $X^{(t)}, t\in \{0,1,...,q-1\}$.
For each element, we have
\begin{equation}  \label{eq:bitDecomposition} \small
    w_{i,j}^{(s)} = (w_{i,j}\gg s)\&1, \quad x_{i,j}^{(t)} = (x_{i,j}\gg t)\&1
\end{equation}
Then, we compute the \texttt{bmma} API for $pq$ times for each combination of $s$ and $t$:
\begin{equation*} \small
    Y^{(s,t)} = \text{\texttt{bmma}}(W^{(s)}, X^{(t)})
\end{equation*}
Finally, we conduct bit combination to generate the $32$-bit output matrix $Y$:
\begin{equation*} \small
    Y = \sum_{s=0}^{p-1}\sum_{t=0}^{q-1} Y^{(s,t)}*2^{s+t}
\end{equation*}

\textbf{Cost Analysis.}
The cost of arbitrary-precision computation comes from three parts: bit decomposition, tensor core computation, and bit combination.
Given a $p$-bit weight matrix and a $q$-bit data matrix of shape $n \times n$, bit decomposition shows complexity of $O((p+q)n^2)$ since we need $O(pn^2)$ operations to split each $p$-bit element from A into $p$ $1$-bit elements and another $O(qn^2)$ operations to split each $q$-bit element from B into $q$ $1$-bit elements. The bit combination shows complexity of $O(pqn^2)$, since we have $pq$ matrices $Y^{(s,t)}$ of shape $n\times n$ and need to add elementwisely.
This overhead is negligible compared with the $O(n^3)$ complexity in the Tensor Core computation.
Note that only 1-bit compute primitives are used for this expensive matrix-matrix multiplication, which significantly reduces the overall latency.

\vspace{-3pt}
\subsection{Data Adaptive Operator Selection} \label{sec:dataAdaptiveOperatorSelection}
While we compute with bit-0 and bit-1 in arbitrary-precision computation, these two values may actually encode diverse values. 
For example, the 1-bit weight matrix in neural networks may encode $-1$ and $1$, instead of $0$ and $1$, in order to improve the accuracy of neural networks.
In this case, bit-0 indicates the value $-1$ and bit-1 indicates the value $1$.
To support this diversity in the encoded data, we introduce \textit{data adaptive operator selection} by adopting different bit operations in Tensor Cores (\textit{i.e.}, \texttt{XOR} and \texttt{AND}).
In particular, we support three cases, where we first conduct bit operations and then accumulate with \texttt{popc} (\textit{i.e.}, population count \cite{popc} that counts the number of set bits).
The \textit{Case-I} is that both $W$ and $X$ encode $0$ and $1$, where we choose logical \texttt{AND} operation.
For example, given a 1-bit vector $W=[0,1]$ and a 1-bit vector $X=[1,1]$, we use \texttt{AND} operation to compute as 
\begin{equation*} \small
    W X = \text{\texttt{popc}}(\text{\texttt{AND}}([0,1], [1,1])) =  \text{\texttt{popc}}([0,1]) = 1
\end{equation*}

The \textit{Case-II} is that both $W$ and $X$ encodes $-1$ and $+1$, where we select logical XOR operation.
For example, given two 1-bit vectors $W=[-1,1]$ and $X=[1,1]$, we first map $-1$ to $0$ and compute as
\begin{equation*} \small
    W X = n - 2*\text{\texttt{popc}}(\text{\texttt{XOR}}([0,1],[1,1])) = n-2*\text{\texttt{popc}}([0,1]) = 0
\end{equation*}
Here, $n$(=2) is the length of the vector.

The \textit{Case-III} is that $W$ encodes $-1$ and $+1$, while $X$ encodes $0$ and $1$.
For example, we may need to compute the multiplication of two 1-bit vectors $W=[-1,1]$ and $X=[1,0]$.
This case happens frequently in neural networks with a 1-bit weight matrix $W$ and a $q$-bit feature matrix $X$ with $q>1$.
In this case, naively adopting \texttt{XOR} or \texttt{AND} does not work, since there are three values $-1$, $0$, and $1$ that cannot be easily encoded with 1 bit.
To this end, we incorporate a linear transformation on $W$ and compute with only $\texttt{AND}$ operation.
Our key observation is that $W$ can be transformed into a vector with only $0$ and $1$ by adding a constant vector $\mathbf{J}_{2} = [1,1]$:
\begin{equation*} \small
    \hat{W} = \frac{W + \mathbf{J}_2}{2} = [0,1]
\end{equation*}
Then, we compute $\hat{W}X=0$ with \texttt{AND} operation as Case-I.
Finally, we recover the value $WX$ by another linear transformation:
\begin{equation*} \small
    WX = 2\hat{W}X - \mathbf{J}_2X = 2*0 - 1 = -1
\end{equation*}
Note that $\mathbf{J}_2$ is a constant vector that can be cached in Tensor Core fragment and does not introduce extra memory overhead.

%% file: 04_APKernel.tex
\section{Arbitrary Precision Layer Design}

In this section, we propose the Arbitrary-Precision Matrix Multiplication (APMM) for fully connected layers and Arbitrary-Precision Convolution (APConv) for convolution layers.

\vspace{-3pt}
\subsection{Arbitrary-Precision Matrix Multiplication} \label{sec:APMM}
Arbitrary-Precision Matrix Multiplication (APMM) takes the decomposed $1$-bit weight matrix $W^{(s)}$, $s\in \{0,...,p-1\}$, the decomposed $1$-bit feature matrix $X^{(t)}, t\in \{0,...,q-1\}$, and computes output matrix $Y = \sum_{s=0}^{p-1}\sum_{t=0}^{q-1} Y^{(s,t)}*2^{s+t}$.
By default, APMM generates $32$-bit output to avoid data overflow for large matrices and match the $32$-bit output in Tensor Core compute primitives.
APMM also supports arbitrary-precision output (\textit{e.g.}, \texttt{int2}) when APMM is used as a hidden layer in neural networks (NNs) and the output is consumed by the next APMM-based NN layer.

\begin{figure*}
    \centering
    \includegraphics[width=0.8\textwidth]{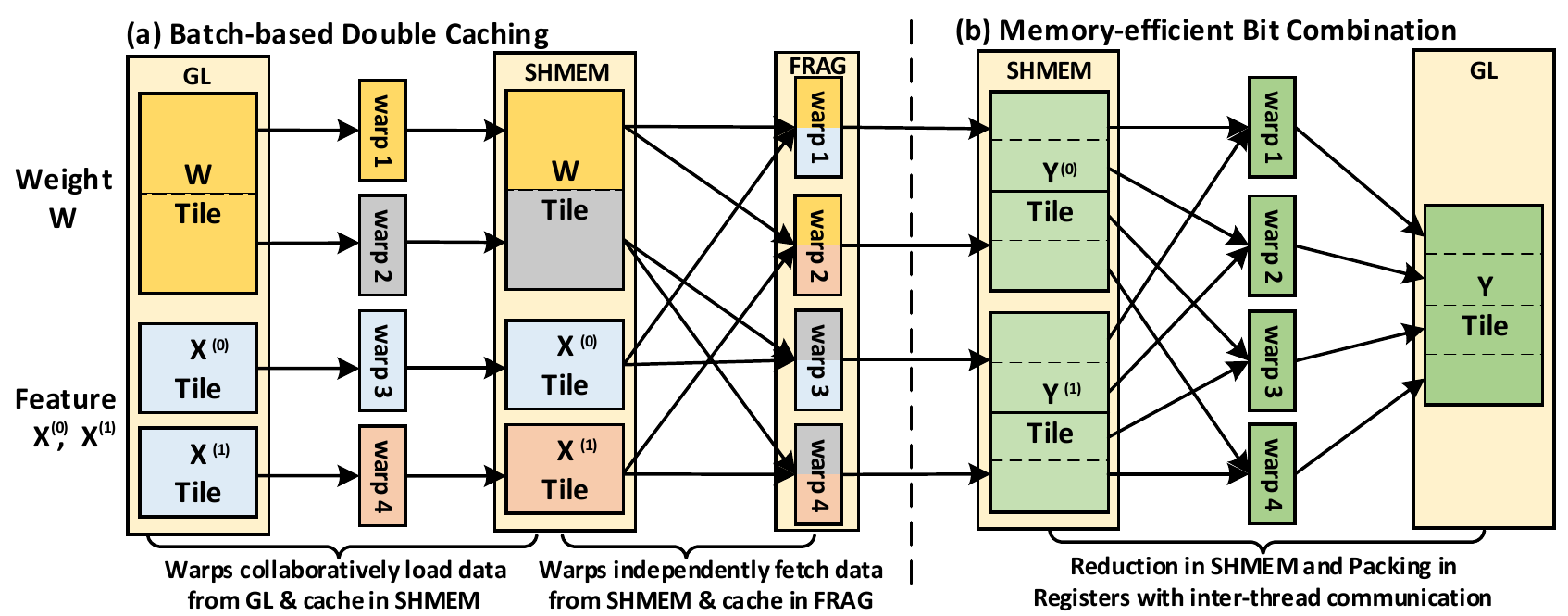}
    \vspace{-7pt}
    \caption{Illustration of APMM. GL: GLobal memory. SHMEM: SHared Memory. FRAG: FRAGment.}
    \vspace{-7pt}
    \label{fig:APMMIllustration}
\end{figure*}

Considering that APMM essentially computes an arbitrary precision GEneral Matrix-Matrix multiplication (GEMM) kernel with multiple Binary Matrix-MAtrix multiplication (BMMA) kernels, one naive strategy is to build upon existing BMMA kernels \cite{TPDS,li2019bstc}.
In particular, we can use existing BMMA kernels to multiply each pair of $W^{(s)}$ and $X^{(t)}$ and accumulate $W^{(s)}X^{(t)}$ to the output matrix $Y$.
However, this approach shows significant inefficiency due to two reasons.
First, this approach ignores the data reuse opportunity since the same weight matrix tile from $W^{(s)}$ can be multiplied with different feature matrix tiles from $X_{t1}$ and $X_{t2}$.
Second, this approach requires extra communication across BMMA kernels, such that reducing $W^{(s)}X^{(t)}$ into $Y$ leads to significant global memory access.
We show our efficient APMM design in Figure \ref{fig:APMMIllustration}.
It includes a \textit{batch-based double caching} to facilitate the data reuse and a \textit{memory-efficient bit combination} to accelerate the accumulation and optionally generate the arbitrary-precision output.
Here, we illustrate the design with $1$-bit $W$ and $2$-bit $X$ for notation simplicity while arbitrary-precision $W$ and $X$ are supported.

\vspace{3pt}
\noindent \textbf{(a) Batch-based Double Caching.}
Batch-based double caching exploits two GPU memory hierarchies (\textit{i.e.}, shared memory and fragment located in registers) to cache matrix tiles and facilitate data reuse in APMM computation, as illustrated in Figure \ref{fig:APMMIllustration}(a).
Considering the limited size of shared memory and fragment, we tile weight matrices $W^{(s)}$ and feature matrices $X^{(t)}$ such that these tiles can be cached in GPU memory hierarchies.
Formally, given $W^{(s)}$ of shape $M\times K$ and $X^{(t)}$ of shape $N\times K$, we first tile $W^{(s)}$ along the $M$ dimension into block matrix tiles of shape $b_m \times b_k$.
Similarly, we tile $X^{(t)}$ along the $N$ dimension into block matrix tiles of shape $b_n \times b_k$.
Here, each GPU block will multiply one pair of block matrix tiles and generate an output matrix tile of shape $b_m \times b_n$.
Considering that Tensor Cores compute at the warp level, we further tile $W^{(s)}$ into warp matrix tiles of shape $w_m\times w_k$ and $X_s$ into $w_n \times w_k$ such that each warp computes an output tile of shape $w_m \times w_n$.
To match with the $8\times8\times 128$ \texttt{bmma} compute primitive of Tensor Cores, each warp will slide along $w_m$, $w_n$, and $K$ dimension during computation.
Note that these tiling sizes have a significant impact on the performance, which will be analyzed in Section \ref{sec:performanceAnalysis}.

Batch-based double caching first adopts a batch strategy to improve inter-thread parallelism and achieve high performance.
Existing works on binary neural networks \cite{li2019bstc,TPDS} report that the GEMM size in NN workload is usually small (\textit{e.g.}, $512\times 512$) and use small matrix tiling sizes (\textit{e.g.}, $32\times 32$) to improve the inter-thread parallelism.
However, this approach leads to low intra-thread parallelism and prevents data reuse.
Instead, our batch strategy virtually transforms multiple small BMMAs into a large BMMA.
In particular, given $W^{(s)}, s\in\{1,...,p-1\}$ of shape $M\times K$ and $X^{(t)}, t\in \{1,...,q-1\}$ of shape $N\times K$, we batch these small matrices into $W_{B}$ of shape $pM \times K$ and $X_B$ of shape $qN \times K$ and compute using a single large BMMA.
Here, we implement a ``virtual" batch strategy during the data loading procedure by dynamically deciding the global memory address of the corresponding matrix tile such that no additional memory movement is involved.

Batch-based double caching then exploits two GPU memory hierarchies to facilitate data reuse at different levels.
The first level is shared memory caching to reuse matrix tiles from $W^{(s)}$ and $X^{(t)}$.
Here, a naive strategy is that each warp independently loads a weight tile and a feature tile for computation.
However, we observe that the same weight tile may be multiplied with feature tiles from different 1-bit feature matrices $X^{(0)}$ and $X^{(1)}$, as illustrated in Figure \ref{fig:APMMIllustration}(a).
To this end, our design requires all warps to first collaboratively load $b_m\times b_k$ weight data and $b_n\times b_k$ feature data from global memory to shared memory.
Then, each warp fetches its own matrix tiles from shared memory.
In this way, we can significantly reduce global memory access by exploiting fast shared memory.


The second level is fragment caching to continuously store output tiles in the same Tensor Core fragment.
Since Tensor Core compute primitives require to accumulate in 32-bit Tensor Core fragments, the output tiles usually consume a large memory space compared with the 1-bit input data.
Moving output tiles between shared memory and Tensor Core fragment may lead to heavy shared memory access.
Moreover, existing dissecting works \cite{jia2019dissecting,jia2018dissecting} reveal that fragment is composed of registers and one GPU block of 8 warps can provide up to 256 KB Fragment, which is much larger than shared memory.
To this end, as iterating through the K dimension during computation, we continuously use multiple fragments to cache output tiles of shape $b_m \times b_n$ for reducing shared memory access and caching more feature and weight tiles in shared memory.

\vspace{1pt}
\noindent \textbf{(b) Memory-efficient Bit Combination.}
Bit combination consumes 32-bit BMMA outputs $Y^{(s,t)} \in \text{\texttt{int32}}^{M\times N}$ and generates $32$-bit APMM outputs $Y\in \text{\texttt{int32}}^{M\times N}$ as $Y = \sum_{s=0}^{p-1}\sum_{t=0}^{q-1} Y^{(s,t)}*2^{s+t}$.
`Bit combination can also generate arbitrary precision output when it is utilized as a NN hidden layer and its output is consumed by the next NN layer.
Overall, bit combination takes only $O(pqMN)$ computation complexity, which is significantly lower than the computation complexity of GEMM operations.
However, there are two potential memory bottlenecks in bit combination, which have a significant performance impact.
The first one is global memory access when reducing $32$-bit BMMA outputs to $32$-bit APMM outputs.
In a naive implementation that independently conducts BMMA and bit combination, bit combination usually introduces similar latency as the BMMA kernel.
The main reason is that, while Tensor Cores provide significantly higher computation throughput than CUDA Cores, the global memory bandwidth remains the same.
The second one is the shared memory access when converting $32$-bit APMM outputs to arbitrary-precision outputs.
In this procedure, we usually need to pack low-bit values (\textit{e.g.}, 2-bit) in registers from different threads to a single memory-aligned value (\textit{e.g.}, 32-bit) before storing to global memory.
Relying on shared memory for data exchange across threads may lead to heavy shared memory access.


Memory-efficient bit combination includes two novel designs to mitigate memory overhead.
The first design includes a semantic-aware workload allocation and an in-shared-memory reduction.
In particular, at the data loading phase of BMMA, we load feature tiles and weight tiles of the same spatial location such that their multiplication outputs can be reduced directly.
As illustrated in Figure \ref{fig:APMMIllustration}, instead of loading a $b_n \times b_k$ feature tile of $X^{(0)}$ or $X^{(1)}$, we load two $0.5b_n \times b_k$ feature tiles of both $X^{(0)}$ and $X^{(1)}$ with the same matrix index.
In this way, we can reduce $WX^{(1)}$ and $WX^{(0)}$ directly in shared memory and mitigate global memory access while not degrading the BMMA performance.


The second design incorporates an element-wise routine and an inter-thread communication to pack low-bit values and mitigate shared memory overhead.
The element-wise routine is a user-defined interface to provide diverse support of quantization and batch normalization across NN layers.
This routine applies to individual 32-bit reduced values in registers.
Given a 32-bit value in a register, this routine may quantize it into a $p$-bit value that is still stored in the 32-bit register with the first $32-p$ bits as zeros.
This routine also includes bit decomposition (Equation \ref{eq:bitDecomposition}) that splits this $p$-bit value in a register to $1$-bit values in $p$ registers.
After that, we use a \texttt{\_\_ballot\_sync} API to enable inter-thread communication and directly pack the $1$-bit values across threads into $32$-bit values that can be stored to the global memory.

\vspace{-3pt}
\subsection{Arbitrary-Precision Convolution (APConv)} \label{sec:APConv}
APConv takes the decomposed 1-bit weight matrix $W^{(s)}$ of shape $C_{out} \times C_{in} \times K\times K$, the decomposed 1-bit feature matrix $X^{(t)}$ of shape $BS \times C_{in} \times Height \times Width$, and generates output matrix $Y$.
Here, $C_{out}$ is the number of output channels, $C_{in}$ is the number of input channels, $K$ is the kernel size, $BS$ is the batch size.
Existing works on bit-level convolution usually adopt a direct convolution design \cite{li2019bstc,TPDS} to improve the GPU utilization.
However, these methods ignore the data reuse opportunity and introduce heavy global memory access.
In addition, APConv on a $p$-bit weight and a $q$-bit feature usually has $pq$ times workload than the BConv on the same weight and feature size, which can easily contribute to high GPU utilization.
To this end, APConv incorporates the batch-based double caching design as APMM to mitigate the global memory access.
However, there are still two key challenges that distinguish APConv from APMM.
The first is the data organization where naively reading the $K\times K$ feature map may easily lead to un-coalesced memory access.
The second is the data padding where simply padding zeros may lead to erroneous results.
To tackle these challenges, we propose \textit{channel-major data organization} and \textit{input-aware padding design}.
\begin{figure}
    \centering
    \includegraphics[width=0.9\linewidth]{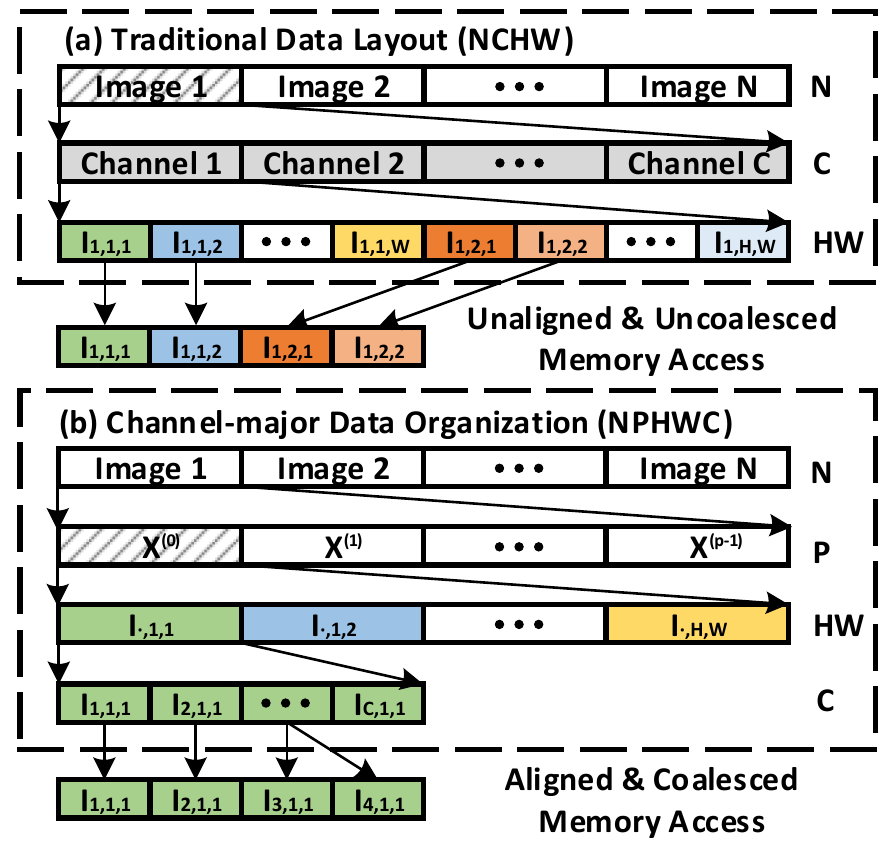}
    \vspace{-7pt}
    \caption{Illustration of Channel Major Data Organization (NPHWC). P indicates the number of bits. $I_{chw}$ indicates the image pixel at the $c$-th channel, $h$-th height, and $w$-th width.}
    \vspace{-10pt}
    \label{fig:channelMajorDataOrganization}
\end{figure}

\vspace{3pt}
\textbf{(a) Channel-Major Data Organization.}
Channel-major data organization transforms un-coalesced and unaligned memory access to a coalesced and aligned one for improving performance.
Traditional data organization for $32$-bit convolution usually employs a NCHW design, as illustrated in Figure \ref{fig:channelMajorDataOrganization}(a).
However, naively borrowing this design to APConv leads to un-aligned and un-coalesced memory access due to two reasons.
First, multiple $P$-bit (\textit{e.g.}, 3-bit) elements usually cannot be packed into an aligned 32-bit element, which is required for valid GPU reads and writes.
Using a 32-bit element to store a $P$-bit element will introduce extra memory overhead.
Second, convolution operations usually read only $K$ continuous elements (or $KP$ bits) due to the $K\times K$ kernel size, which may lead to un-coalesced memory access. 

We design a channel-major data organization as illustrated in Figure \ref{fig:channelMajorDataOrganization}(b).
There are two key design choices.
First, we split a $P$-bit feature matrix into $P$ 1-bit feature matrices and store each 1-bit feature matrix consecutively.
In this way, we can provide aligned memory access for each 1-bit feature matrix and support arbitrary precision $P$.
Second, we consecutively store all channels of elements with the same spatial location.
Since convolution layers usually have $128C, C\in \mathbb{N}$ channels, this usually leads to coalesced memory access during computation.

\vspace{3pt}
\textbf{(b) Input-aware Padding Design.}
Input-aware padding design adaptively adjusts padding values according to input values.
As mentioned in Section \ref{sec:dataAdaptiveOperatorSelection}, when the weight W encodes $-1$ and $1$ with $0$ and $1$, we cannot naively padding $0$ since $0$ represents $-1$.

We propose three padding strategies according to the input data.
First, when both weight and feature encode $0$ and $1$, we simply pad zeros for features.
In this case, padding $0$ for features will only add extra $0$'s for arbitrary weight values, which does not change the computation result.
Second, when both weight and feature encode $-1$ and $1$, we pad $1$ for features and use an extra \texttt{counter} flag to track the number of $0$'s when the convolution weight moves outside the input image frame.
We will subtract \texttt{counter} to amend the corresponding convolution results.
Third, when weight encodes $-1$ and $1$ and feature encodes $0$ and $1$, we pad $0$ to features and do not change the convolution results.

\vspace{-7pt}
\subsection{Performance Analysis} \label{sec:performanceAnalysis}
In our \Mname~kernel design, there are six tuning knobs -- the block tiling sizes $b_m$, $b_n$, $b_k$, and the warp tiling sizes $w_m$, $w_n$, $w_k$.
These tiling sizes bring a trade-off between the Thread-Level Parallelism (TLP) and the Instruction Level Parallelism (ILP), especially the compute intensity (CI).
Here, we focus on block tiling sizes, since we empirically observe that utilizing $8$ warps per block and splitting the block workload evenly across warps provide the best performance (\textit{i.e.}, $w_m = b_m/4$, $w_n = b_n/2$, $w_k=b_k$).
In this subsection, we first analyze the performance impact of individual tuning knobs.
Then, we propose an autotuning strategy to maximize the performance.
Since APMM and APConv share the same batch-based double caching strategy, we use the same autotuning strategy for these two kernels.

\vspace{-7pt}
\subsubsection{Performance Model}
TLP refers to the thread-level parallelism in terms of the number of threads in use.
Intuitively, larger TLP can improve GPU utilization and kernel performance \cite{li2015transit, li2016x}.
Formally, given a p-bit weight matrix of shape $M\times K$, a q-bit feature matrix of shape $K\times N$ and the matrix tiling size $b_m\times b_n$, we define the TLP as
\begin{equation} \label{eq:TLP} \small
TLP=
\frac{pM \times qN}{b_m \times b_n}
\end{equation}
We ignore the number of threads for each block since it is a constant in our evaluation.
Intuitively, smaller $b_m\times b_n$ may improve TLP, which suggests a small $b_m\times b_n$ especially for small matrices.

Compute intensity (CI) refers to the ratio of computation over memory access on each thread block.
We aim to improve CI for two reasons.
First, a higher CI indicates less memory access and better performance.
While the amount of computation remains the same, the amount of memory access may be reduced significantly by data reusing and hyper-parameter tuning.
Second, a higher CI on a thread block provides more opportunities for latency hiding.
Formally, for a matrix tile, we compute the amount of global memory access as $b_m \times b_k + b_n \times b_k$ when reading a $b_m\times b_k$ weight tile and a $b_m\times b_k$ feature tile.
We compute the amount of computation as $2\times b_m \times b_n \times b_k$ from the matrix-matrix multiplication.
Finally, we compute CI as
\begin{equation} \label{eq:CI}\small
    CI = \frac{2\times b_m \times b_n}{b_m + b_n}
\end{equation}
Note that CI can be increased when $b_m$ and $b_n$ are increased.
We also observe that CI is independent of $b_k$ such that we can use smaller $b_k$ to leave space for larger $b_m$ and $b_n$, especially when the shared memory size is a limiting factor.
In our evaluation, we fix $b_k$ as $128$ by default.


\subsubsection{Auto-tuning}

During APNN-TC kernel design, there is a large search space on the complex interaction between matrix size ($M$, $N$, and $K$), weight bit $p$, feature bit $q$, and block tiling size $b_m$ and $b_n$. Note that the selected parameters may also be different on various GPUs according to computation and memory capabilities.
To this end, we propose a heuristic algorithm to provide a faster search procedure in this large search space.
Formally, given the matrix size $M$, $N$, $K$, the weight bit $p$, the feature bit $q$, the algorithm selects $b_m, b_n\in \{16,32,64,128\}$ in two steps.
First, we compute the TLP of each combination of $b_m$ and $b_n$.
We put these combinations in a priority queue, where a higher TLP leads to a high priority.
Second, we pop individual combinations in the priority queue.
We stick to the first combination with the highest TLP if its TLP is already smaller than a threshold $T$.
Otherwise, we continuously pop and select combinations in the priority queue to improve CI while ensuring TLP is larger than $T$.
We empirically set $T$ as $64$ in our evaluation.
Note that different block tiling sizes share the same data layout such that there is no overhead when consecutively executing two layers with different block tiling sizes.

%% file: 05_APNN.tex
\section{Arbitrary Precision Neural Network Design}
In this section, we introduce our Arbitrary Precision Neural Network (APNN) design.
We first introduce a minimal-traffic dataflow on supporting various precisions across layers in APNN.
Then, we incorporate a semantic-aware kernel fusion to minimize the memory access across layers.

\vspace{-3pt}
\subsection{Minimal-Traffic Dataflow}
Given an \texttt{int8} RGB image, APNN computes a sequence of NN layers with $p$-bit weights and $q$-bit activations and finally generates an \texttt{int32} output logits.
Here, all intermediate layers compute at arbitrary precision by taking a $p$-bit weights and $q$-bit activations and generate $32$-bit outputs.
Note that the \texttt{int1} Tensor Core compute primitive can only generate $int32$ outputs and an extra quantization layer is required to quantizing into $q$-bit activations for the next layer.
For performance consideration, during the initialization of an APNN, we quantize all weights before the model inference computation.
To effectively maintain and transfer arbitrary-bit data, we pack the data bit-by-bit for both weight and feature map, following the data organization discussed in Section \ref{sec:APConv}.

The input layer and the output layer have different precisions from the intermediate layers.
As is the common practice with \texttt{int8} image inputs, the input layer requires an extra quantization layer that quantizes $8$-bit inputs into $q$-bit activations.
The output of the input layer will also be the quantized arbitrary-bit feature map serving as the input for the following intermediate layers. 
In the output layer, Tensor Core computation results will be directly used for the final softmax logits computation.
Thus, we do not apply quantization after the output layer.


%



%
%

\vspace{-3pt}
\subsection{Semantic-aware Kernel Fusion}
Besides APMM and APConv discussed previously, there are still multiple important layers in APNN, including quantization, Batch Normalization (BN), pooling, and ReLU.
Given all scalars $x_{i,j}$ in the $i^{th}$ layer, quantization element-wisely converts \texttt{int32} values $x_{i,j}$ to \texttt{q}-bit values $y_{i,j}$:
\begin{equation*} \small
    y_{i,j} = \lfloor(x_{i,j} - z_i)/s_i\rfloor
\end{equation*}
Here, $z_i$ is a $32$-bit scalar zero-point, $s_i$ is the scaling scalar, and $\lfloor\cdot \rfloor$ is the floor function.
BN \cite{ioffe2015batch} is another major component in NNs for tackling the covariate shift problem and facilitating NN training:
\begin{equation} \small \label{eq: fuse BN}
    y_{i,j} = \frac{x_{i,j} - \mathbb{E}[x_{i,*}]}{\sqrt{Var[x_{i,*} + \epsilon]}}\cdot\gamma_j + \beta_j
\end{equation}
where $\mathbb{E}$ and $Var$ are expectation and variance across the batch, $\gamma_j$ and $\beta_j$ are two learned parameters.
Pooling splits the feature map spatially into $k\times k$ grids and generates 1 scalar output for each grid by computing the average or the maximum value in each grid.
ReLU takes individual input values $x_{i,j}$ and generates output values $y_{i,j} = max(x_{i,j},0)$.

While these operations have linear time complexity to the size of feature maps and consume significantly less computation than APConv and APMM kernels, these operations may still introduce heavy latency due to the expensive memory access.
Indeed, while Tensor Cores provides significantly improved computation capability, Tensor Cores share the same memory bandwidth with CUDA Cores on GPUs.
Moreover, we observe that these values are usually computed element-wisely and do not require heavy communication across GPU threads. We propose a semantic-aware kernel fusion to minimize memory access.
We first fuse APMM/APConv with its following quantization, BN, pooling, and ReLU kernels into a single kernel to minimize the global memory access.
In particular, these following layers can be seamlessly applied once the convolution results become available at the shared memory. 
This can improve the computation intensity for individual convolution kernels meanwhile reducing the global memory access from invoking an additional batch normalization kernel.
Second, considering that these following layers usually compute at scalar level, we can further reduce shared memory access by directly reusing values in registers \cite{li2016critical}.
For example, when a APMM layer is followed by a BN layer, a quantization layer, and a ReLU layer, we directly compute the output scalar as
\begin{equation*}
    \lfloor max(\frac{x_{i,j} - \mathbb{E}[x_{i,*}]}{\sqrt{Var[x_{i,*} + \epsilon]}}\cdot\gamma_j + \beta_j - z_{i}, 0)/s_i \rfloor
\end{equation*}
Note that we only need to load a scalar once to a register and avoids unnecessary shared memory access.

%% file: 06_evaluation.tex
\section{Evaluation}
In this section, we evaluate APNN-TC under diverse precisions and show the benefits of arbitrary-precision computation in performance and accuracy.
\begin{figure*}[t]
    \begin{subfigure}{0.48\textwidth}
        \centering
        \includegraphics[width=\linewidth,height=5cm]{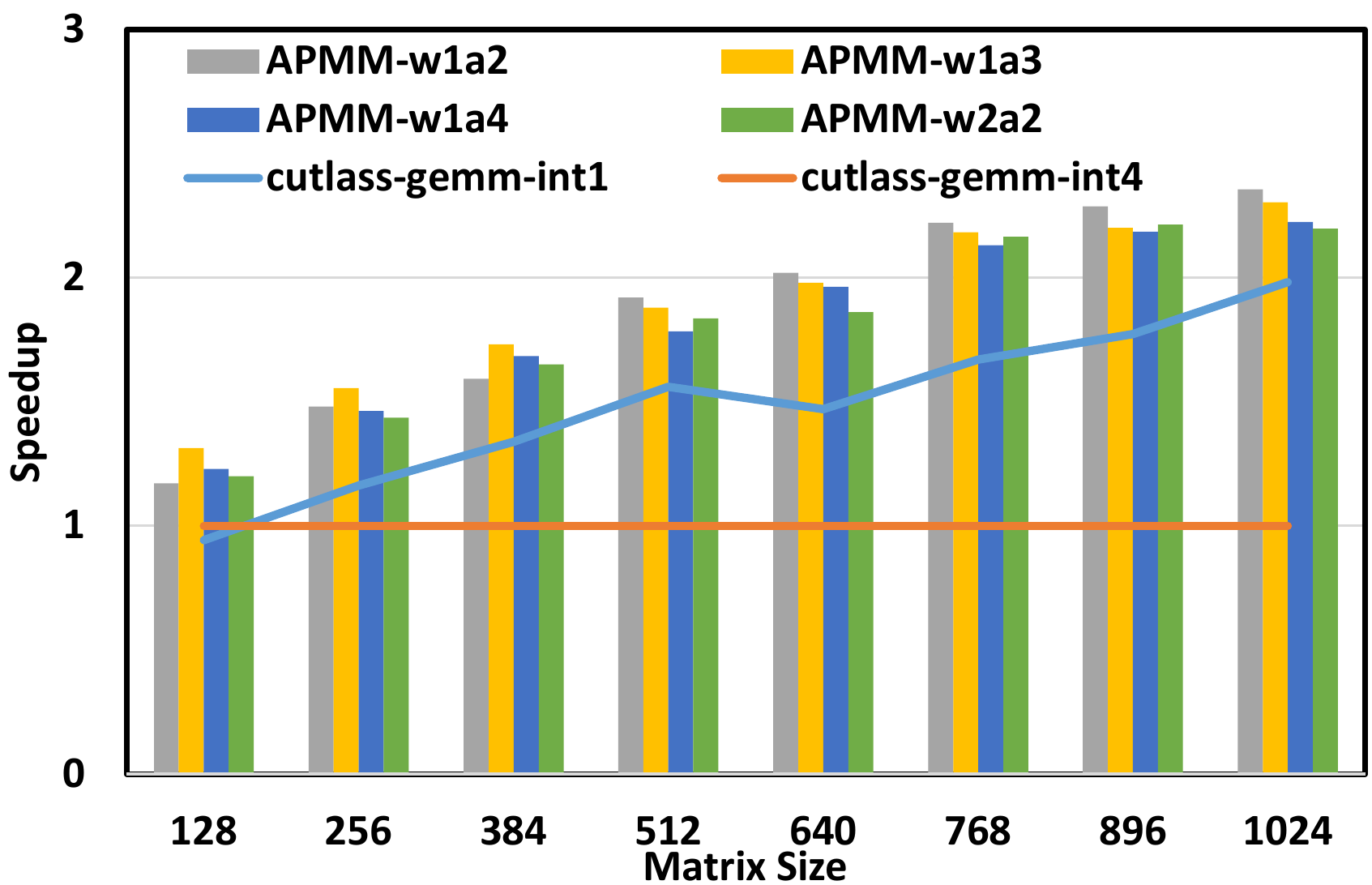}
        \vspace{-10pt}
        \caption{Over CUTLASS-GEMM-INT4}
        \vspace{-10pt}
    \end{subfigure}
    \hfill
    \begin{subfigure}{0.48\textwidth}
        \centering
        \includegraphics[width=\linewidth,height=5cm,trim=0 0.25cm 0 -0.25cm]{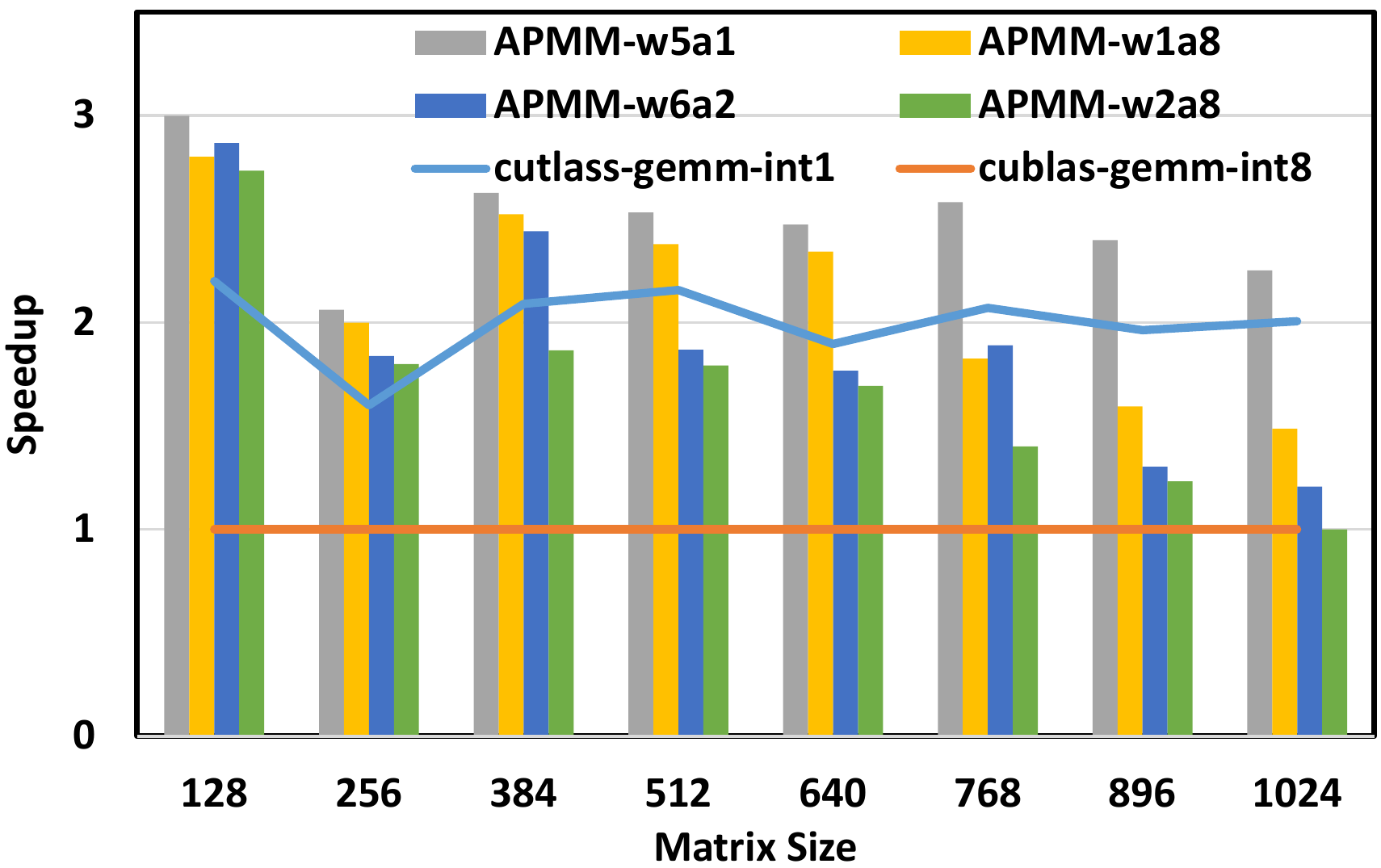}
        \vspace{-10pt}
        \caption{Over CUBLAS-GEMM-INT8.}
        \vspace{-10pt}
    \end{subfigure}
    \caption{APMM Performance on RTX 3090.}
    \vspace{-5pt}
    \label{fig:APMMOverall-3090}
\end{figure*}
\begin{figure*}[t]
    \begin{subfigure}{0.48\textwidth}
        \centering
        \includegraphics[width=\linewidth,height=5cm]{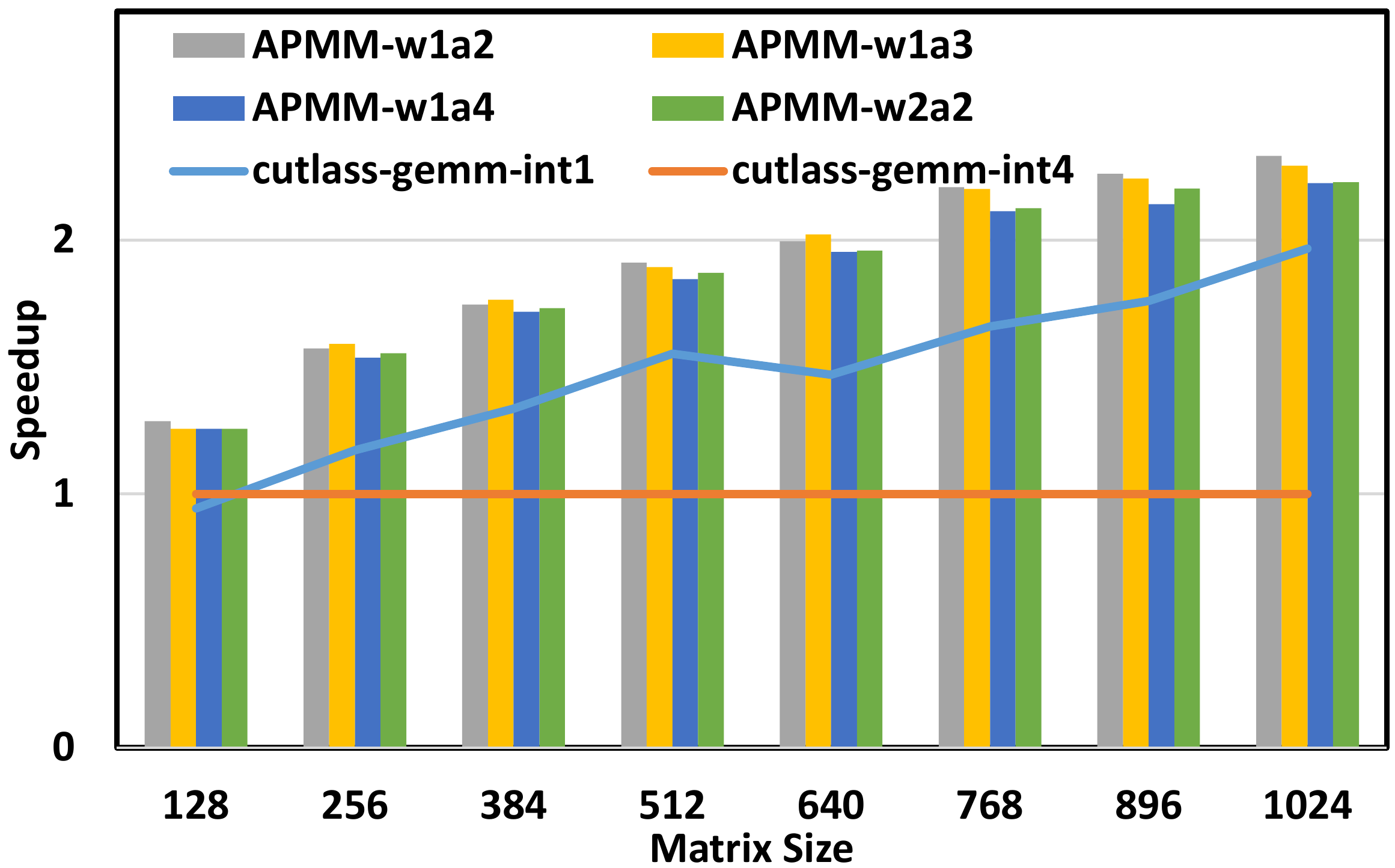}
        \vspace{-10pt}
        \caption{Over CUTLASS-GEMM-INT4.}
        \vspace{-10pt}
    \end{subfigure}
    \hfill
    \begin{subfigure}{0.48\textwidth}
        \centering
        \includegraphics[width=\linewidth,height=5cm]{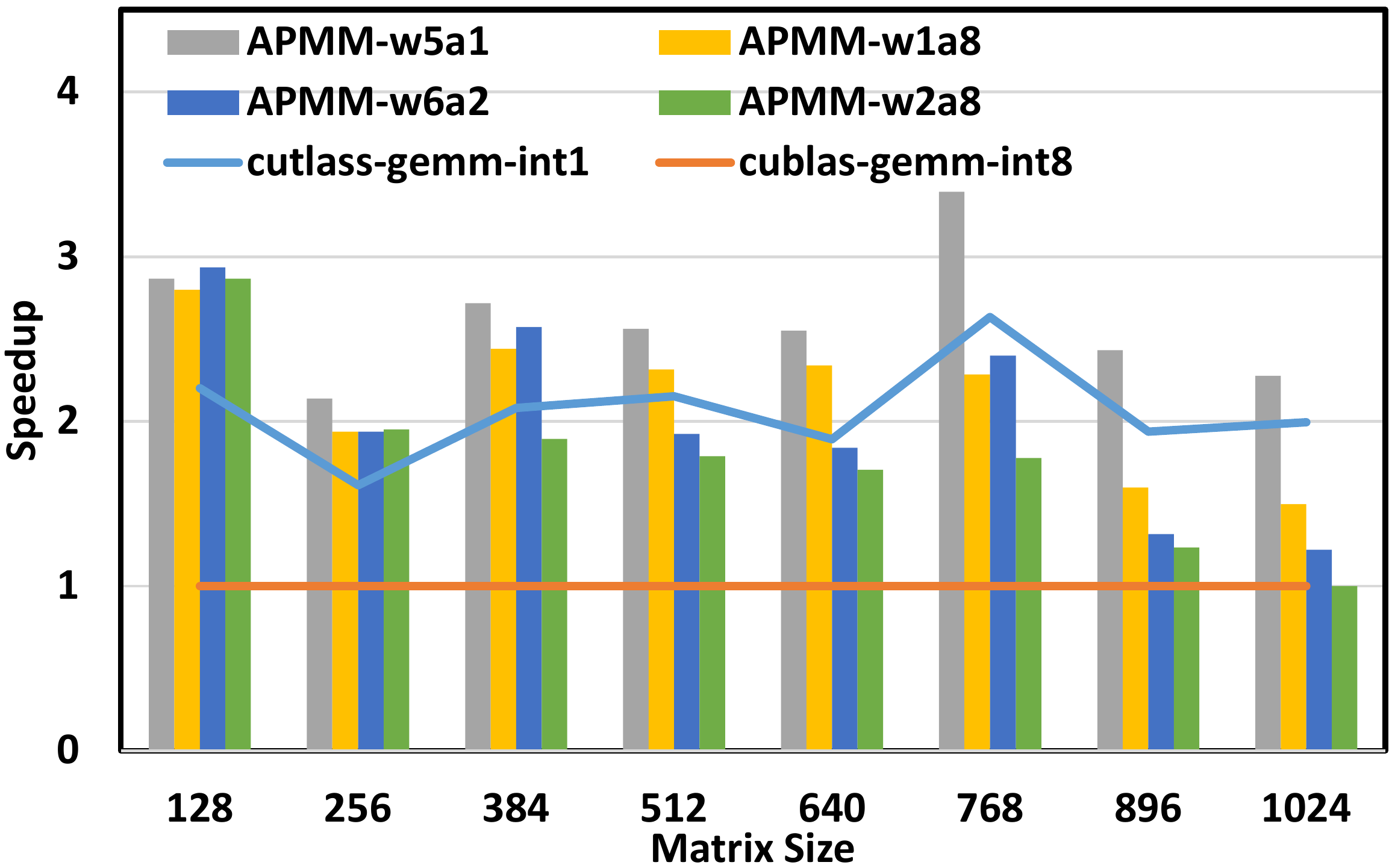}
        \vspace{-10pt}
        \caption{Over CUBLAS-GEMM-INT8.}
        \vspace{-10pt}
    \end{subfigure}
    \caption{APMM Performance on A100.}
    \vspace{-10pt}
    \label{fig:APMMOverall}
\end{figure*}
We evaluate on both Nvidia RTX 3090 and Nvidia Tesla A100.
The RTX3090 GPU is in a ubuntu 16.04 system with Intel(R) Xeon(R) Silver 4110 CPU @ 2.10GHz, 64 GB DDR3 DRAM, gcc-7.5.0, and using CUDA-11.1, CUTLASS-2.5, and CUBLAS-11.1.
The A100 GPU is in a Linux 3.10.0 system with AMD EPYC 7742 64-core CPU, 1TB DDR4, gcc-9.1.0, and using CUDA-11.1, CUTLASS-2.5, and cuBLAS-11.3.
All results reported are the average of 200 times execution.


\vspace{-3pt}
\subsection{APLayer Evaluation}

\subsubsection{APMM Performance} \label{sec:APMMPerformance}
We compare our APMM designs with NVIDIA implementations of low-bit gemm (\textit{i.e.}, \texttt{int1}, \texttt{int4}, and \texttt{int8}) that are accelerated by Tensor Cores.
For \texttt{int8}, we compare with cublas implementation, namely cublass-gemm-int8.
Since \texttt{int1} and \texttt{int4} are not supported in cublas, we compare with cutlass implementation, namely cutlass-gemm-int1 and cutlass-gemm-int4.
Following popular settings in NNs, we compute matrix multiplication of a matrix with shape $B\times K$ and a matrix with shape $K \times N$, where $B=64$ is a popular batch size and $K=N \in \{128, 256, ..., 1024\}$ covers typical fully connected layer dimensions.
According to the precision of our APMM kernel, we name it APMM-wxay, where x indicates the weight bit and y indicates the activation bit.
For example, APMM-w1a2 indicates 1-bit weights and 2-bit activations.
While our APMM is general to support arbitrary precision, we show 8 popular bit combinations due to page limits.
If both weight bits and activation bits are less than $4$ (\textit{e.g.}, w1a2, w1a3, w1a4, w2a2), we compare it against cutlass-gemm-int4.
If either weight bits or activation bits are larger than $4$, we compare it against cublas-gemm-int8.
For each matrix size, we show a speedup of cutlass-gemm-int1 against cutlass-gemm-int4 and cublas-gemm-int8 as the performance benefit when sticking to binary neural networks \cite{li2019bstc, TPDS}.
Since Tensor Core compute primitive supports only $32$-bit outputs, all gemm kernels take low-bit input (\textit{e.g.}, \texttt{int1}, \texttt{int4}, and \texttt{int8}) and generate $32$-bit outputs.

\begin{figure*}[t] \small
    \begin{subfigure}{0.48\textwidth}
        \centering
        \includegraphics[width=\linewidth]{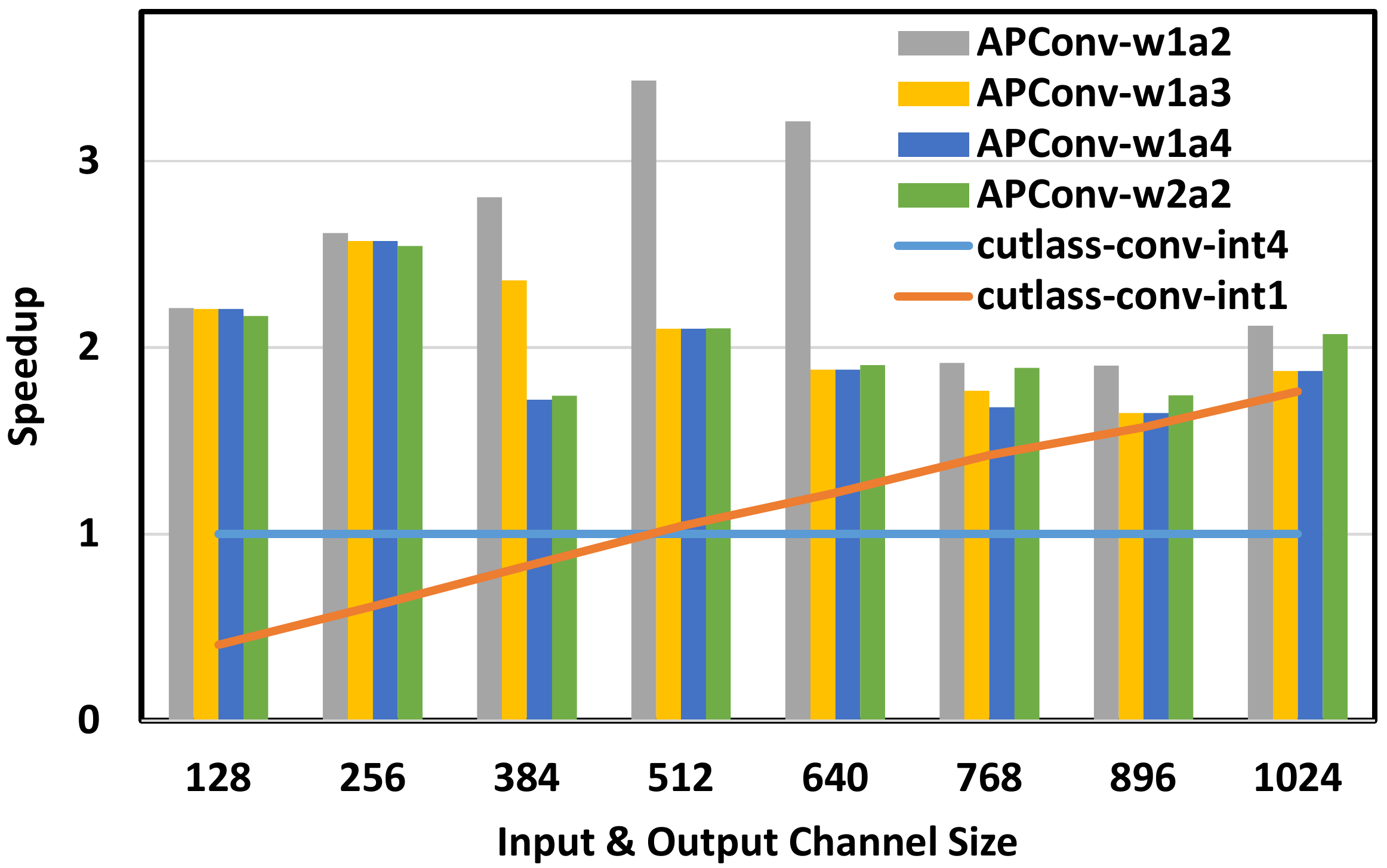}
        \vspace{-10pt}
        \caption{Over CUTLASS-Conv-INT4.}
        \vspace{-10pt}
    \end{subfigure}
    \hfill
    \begin{subfigure}{0.48\textwidth}
        \centering
        \includegraphics[width=\linewidth]{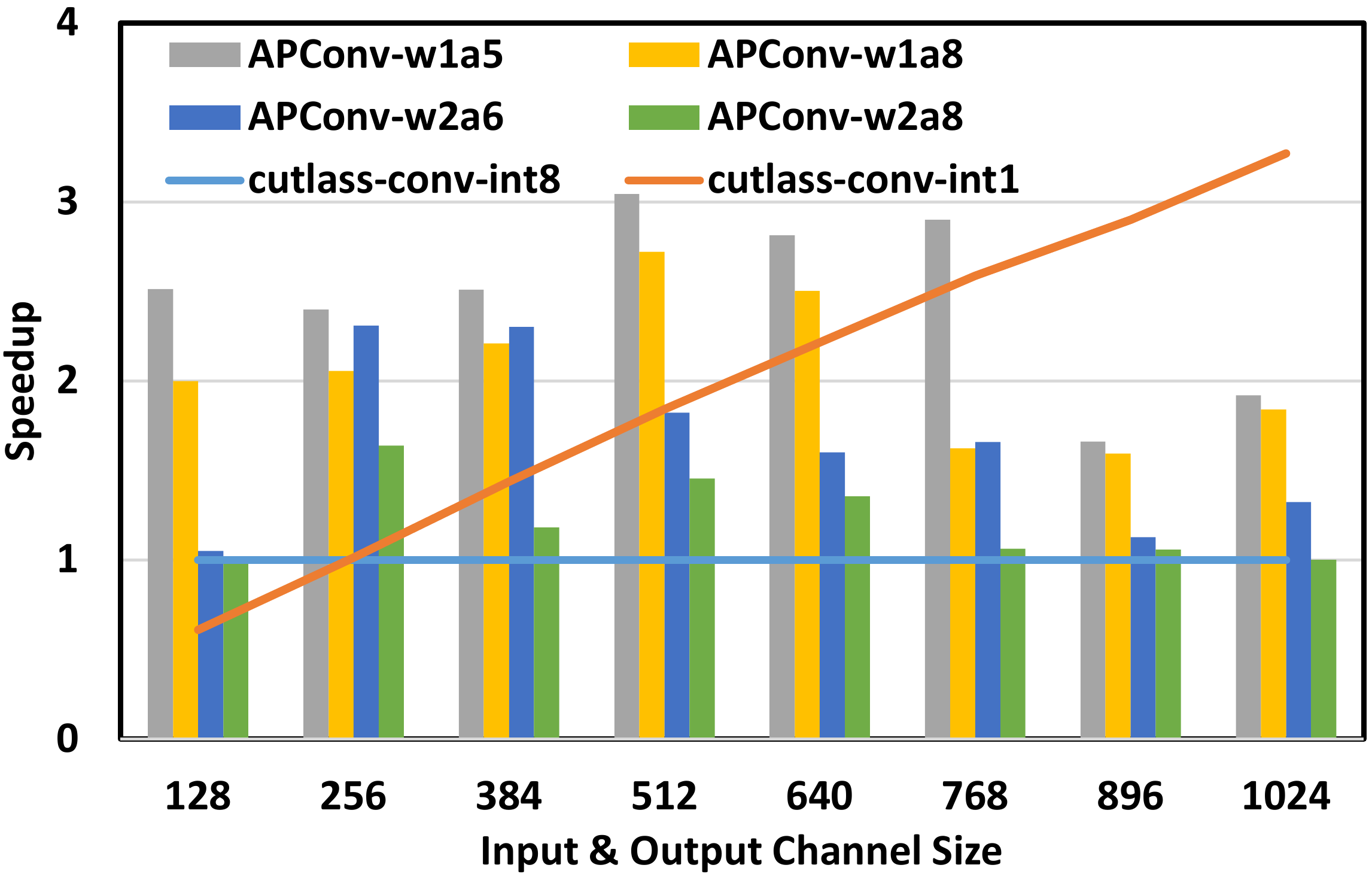}
        \vspace{-10pt}
        \caption{Over CUTLASS-Conv-INT8.}
        \vspace{-10pt}
    \end{subfigure}
    \caption{APConv Performance on RTX 3090.}
    \vspace{-10pt}
    \label{fig:APConv-3090}
\end{figure*}

\begin{figure*}[t] \small
    \begin{subfigure}{0.48\textwidth}
        \centering
        \includegraphics[width=\linewidth,height=5cm]{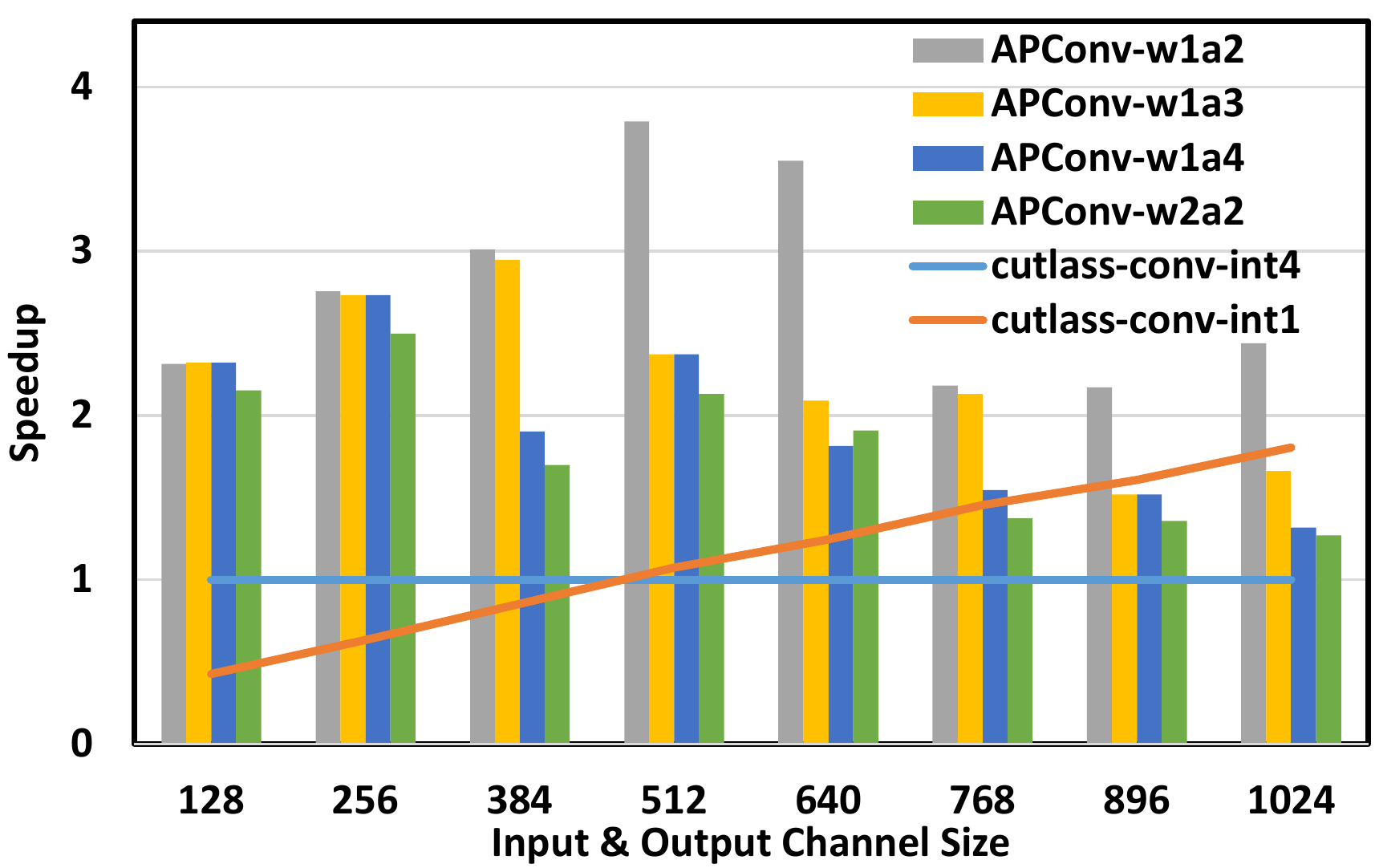}
        \vspace{-10pt}
        \caption{Over CUTLASS-Conv-INT4.}
        \vspace{-10pt}
    \end{subfigure}
    \hfill
    \begin{subfigure}{0.48\textwidth}
        \centering
        \includegraphics[width=\linewidth,height=5cm]{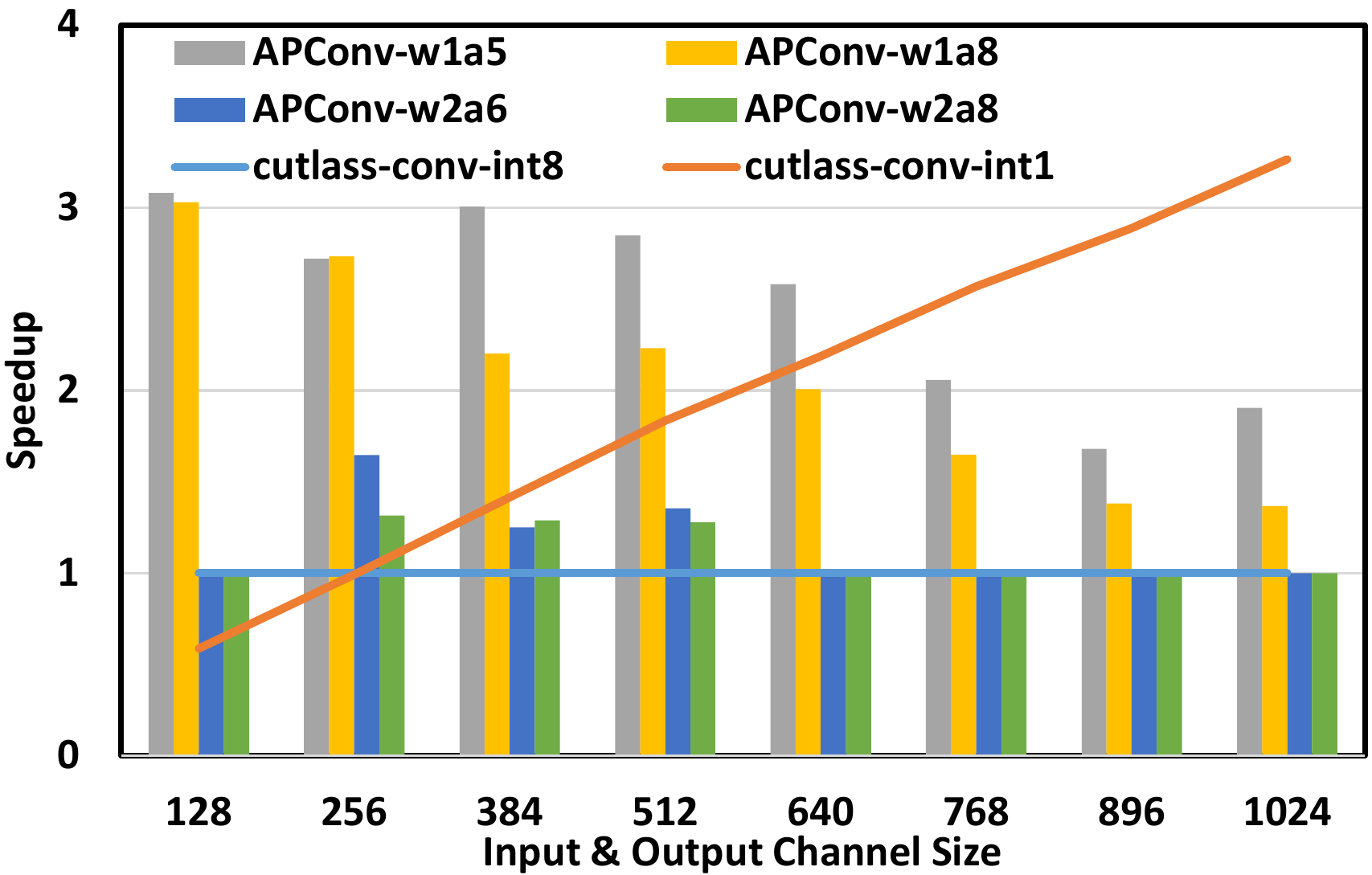}
        \vspace{-10pt}
        \caption{Over CUTLASS-Conv-INT8.}
        \vspace{-10pt}
    \end{subfigure}
    \caption{APConv Performance on A100.}
    \vspace{-10pt}
    \label{fig:APConv-A100}
\end{figure*}

Figure \ref{fig:APMMOverall-3090} shows the results of APMM on RTX 3090.
We compare APMM with cutlass-gemm-int4 in Figure \ref{fig:APMMOverall-3090}(a) and cublas-gemm-int8 in Figure \ref{fig:APMMOverall-3090}(b).
Overall, we have three major observations.
First, APMM can usually achieve significant speedup over baselines.
For example, APMM-w1a2 can achieve up to $2.35\times$ speedup over cutlass-gemm-int4, while APMM-w5a1 can achieve up to $3\times$ speedup over cublas-gemm-int8.
This result demonstrates the performance benefits of emulating arbitrary-precision with \texttt{int1} compute primitives over sticking to \texttt{int4} or \texttt{int8} compute primitives.
Second, APMMs with various weight and activation bits usually show similar performance on small matrices.
For example, APMM-w1a2, APMM-w1a3, APMM-w1a4, and APMM-w2a2 achieves almost the same speedup when N=128 and N=256, even if these kernels have different computation overhead (\textit{e.g.}, $2\times$ from APMM-w1a2 and $4\times$ from APMM-w2a2).
This benefit comes from our batch-based double caching (Section \ref{sec:APMM}(a)), where individual small BMMAs are batched into a large BMMA and computed simultaneously.
Surprisingly, our arbitrary precision computation can even outperform cutlass-gemm-int1 in such cases due to the improved GPU utilization.
Third, we observe a smaller speedup over cublas-gemm-int8 on large matrix sizes, when peak \texttt{int1} performance is achieved.
Our investigation shows that, on RTX 3090, cutlass-gemm-int1 is only $5.9\times$ faster than cublas-gemm-int8, such that emulation is slower than built-in int8 compute primitives on large matrices when peak int1 performance is achieved (\textit{e.g.}, $64\times1024\times1024$ for APMM-w2a8).
We argue that NN workload can still benefit significantly from our APMM since the fully connected layers in NNs usually have small matrix sizes (\textit{e.g.}, $1\times512\times512$ in ResNet-18).
We also show the results of APMM on A100 in Figure \ref{fig:APMMOverall} with similar observations.

\vspace{-5pt}
\subsubsection{APConv Performance} \label{sec:APConvPerformance}
We compare APConv designs with NVIDIA implementations of low-bit convolution that are accelerated by Tensor Cores.
Since cublas does not support \texttt{int1}, \texttt{int4}, AND \texttt{int8} convolution, we use kernels from cutlass.
We name these kernels as cutlass-conv-int1, cutlass-conv-int4, and cutlass-conv-int8.
Similar to APMM, we evaluate 8 types of precision with the name APConv-wxay.
Since convolution kernels have much more hyperparameters than matrix-multiplication kernels, we show the performance under various input and output channels while fixing the input size as $16$ (medium feature size), filter size as $3$ (most frequently used), stride as $1$ (most frequently used), and batch as $1$ (for inference).
Figure \ref{fig:APConv-3090} and \ref{fig:APConv-A100} show the speedup of APConv on RTX 3090 and A100, respectively. APConv can achieve $3.78\times$ speedup over cutlass-conv-int4 and $3.08\times$ speedup over cutlass-conv-int8.
This result shows the significant performance benefit from emulating arbitrary precision with \texttt{int1} over utilizing \texttt{int4} or \texttt{int8}.
Similar to APMM, we also observe a smaller speedup over cutlass-conv-int8 on larges channels due to the limitation of peak int1 performance.
Since RTX3090 and A100 provide similar performance, we will focus on RTX3090 in the following evaluations.

\vspace{-3pt}
\subsection{APNN Evaluation}
In this section, we evaluate the overall APNN performance on three  mainstream neural network models with ImageNet dataset. The details of our evaluated NN models and their corresponding binarized neural network, low-bit (1-bit weight with 2-bit activation), single-precision accuracy precision are listed in Table~\ref{table: CNN models}. 

\begin{table}[t]
\caption{\small APNN Evaluation Setting. We list the dataset, network, input size, output size, and the model accuracy under precisions of BNN (\textit{i.e.}, int1), w1a2 (\textit{i.e.}, 1-bit weights with 2-bit activations), and single-precision floating point.}
\vspace{-5pt}
\scalebox{0.66}{
\begin{tabular}{|c|c|l|c|c|c|c|c|c|c|}
\hline
\textbf{Dataset} &
  \textbf{Network} &
  \textbf{Input Size} &
  \textbf{Output Size} &
  \textbf{Binary} &
  \textbf{w1a2} &
  \textbf{Single} \\ \hline
%
%
%
ImageNet &  AlexNet   \cite{krizhevsky2012imagenet} 
& 224x224x3 & 1000 & 46.1\% & 55.7\% & 57.0\%\\ \hline
ImageNet & VGG-Variant  \cite{cai2017deep}   
& 224x224x3 & 1000 & 53.4\% &  68.8\%  &  69.8\% \\ \hline
%
ImageNet & ResNet-18  \cite{he2016deep} 
& 224x224x3 & 1000 & 51.2\% & 62.6\% & 69.6\%\\ \hline
\end{tabular}}
\label{table: CNN models}
\vspace{-20pt}
\end{table}




\begin{table}[t] \small
\caption{\small APNN Inference Performance on NVIDIA Ampere RTX3090 GPU. Note that latency is measured under a batch of 8 images, throughput is measured under a batch of 128.}
\vspace{-7pt}
\scalebox{0.63}{
\begin{tabular}{|c|r|r|r|r|r|r|}
\hline
\multicolumn{1}{|l|}{\textbf{}} &
  \multicolumn{2}{c|}{\textbf{ImageNet-AlexNet}} &
  \multicolumn{2}{c|}{\textbf{ImageNet-VGG\_Variant}} &
  \multicolumn{2}{c|}{\textbf{ImageNet-ResNet18}} \\ \hline
\textbf{Schemes} &
  \textbf{8 Latency} & \textbf{Throughput} &
  \textbf{8 Latency} & \textbf{Throughput} &
  \textbf{8 Latency} & \textbf{Throughput} 
\\ \hline
\textbf{CUTLASS-Single} 
& 4.43ms  & 2.89$\times10^4$fps
& 25.24ms & 3.89$\times10^2$fps
& 60.96ms  & 1.51$\times10^2$fps\\ \hline

\textbf{CUTLASS-Half-TC}      
& 3.79ms & 3.38$\times10^4$fps
& 24.19ms & 4.67$\times10^2$fps 
& 57.33ms & 1.89$\times10^3$fps \\ \hline

\textbf{CUTLASS-INT8-TC} 
& 13.10ms   & 9.77$\times10^3$fps 
& 25.77ms   & 6.52$\times10^2$fps 
& 57.09ms     & 2.85$\times10^3$fps \\ \hline

\textbf{BNN} 
& 0.69ms & 1.37$\times10^4$fps
& 2.17ms & 3.91$\times10^3$fps
& 0.68ms & 1.89$\times10^4$fps\\ 
\hline 
\hline
\textbf{APNN-w1a2} 
& 0.36ms & 2.85$\times10^4$fps 
& 1.66ms & 5.32$\times10^3$fps 
& 0.64ms & 1.70$\times10^4$fps \\ 
\hline
\end{tabular}}
\label{tbl: APNN Inference Performance on NVIDIA Ampere RTX3090 GPU}
\vspace{-2pt}
\end{table}
We consider two types of configurations for evaluation. 
In the first setting, we focus on a specific low-bit configuration (1-bit weights and 2-bit activations, \textit{i.e.}, w1a2) across different neural network models.
We choose several baselines including neural networks built with single-precision floating-point implementation from CUTLASS~\cite{cutlass} running on CUDA Cores, half-precision implementation from CUTLASS running on Tensor Cores,  INT8 precision implementation from CUTLASS running on Tensor Cores, and the 1-bit binarized neural network running on Tensor Cores based on the state-of-the-art design from~\cite{TPDS}.
As shown in Table~\ref{tbl: APNN Inference Performance on NVIDIA Ampere RTX3090 GPU}, our APNN design running on Tensor Cores can achieve a significant speedup compared with CUTLASS INT8, half and single precision implementations.
This indicates the practical usage of our APNN design in latency-sensitive applications.
Meanwhile, on large batch sizes for throughput performance evaluation, our APNN design also demonstrates its high throughput advantage over these ``standardized'' bit (\textit{e.g.}, 8-bit and half) precision baselines. 
Compared with the 1-bit binarized neural network running on Tensor Cores, our APNN design would demonstrate its significant accuracy improvement (an average 11.67\%) as listed in Table~\ref{table: CNN models}. This can demonstrate the application of our APNN design in some application settings, where the BNN model accuracy performance fails to meet the demands. 
Overall, from the study, we can see that using our APNN design for arbitrary-bit precision computation is a potential way for balancing NN model accuracy and runtime performance.  

\begin{table}[t] \small
    \centering
    \vspace{-5pt}
    \caption{Case Study: APNN of VGG on ImageNet.}
    \vspace{-10pt}
    \scalebox{0.8}{
    \begin{tabular}{|c|c|c|}
    \hline
        \textbf{Scheme} & \textbf{8 Latency (ms)} & \textbf{Throughput (fps)} \\
    \hline
        Float &  25.24 & 3.89$\times10^2$\\
    \hline
        Half  & 24.19& 4.66$\times10^2$\\
    \hline
        INT8  & 25.77 & 6.52$\times10^2$ \\
    \hline
        BNN  &  2.17 & 3.91$\times10^3$\\
    \hline
    \hline
        APNN-w1a2 & 1.66 & 5.32$\times10^3$\\
    \hline
        APNN-w2a2 & 3.08 & 2.59$\times10^3$\\
    \hline        
        APNN-w2a8 & 14.14 & 5.65$\times10^2$\\
    \hline        
    \end{tabular}}
    \label{tab:APNN-various-precision}
    \vspace{-10pt}
\end{table}

In the second setting, we shift our focus towards model runtime performance tradeoff on the VGG network. 
We select several low-bit settings for comparison, including the 1-bit weight with 2-bit activation, 2-bit weight with 2-bit activation, and 2-bit weight with 8-bit activation.
As shown in Table \ref{tab:APNN-various-precision}, APNN-TC significantly reduces latency and improves throughput for w1a2 and w2a2 than INT8 which shows that APNN-TC can bring benefits for many arbitrary-precision computations.
Comparing with INT8, APNN-TC with w2a8 shows lower throughput since we need to compute 16 (=2*8) 1-bit matrices to emulate arbitrary-precision computation, which require more computation than w1a2 with 2 1-bit matrices and w2a2 with 4 1-bit matrices.
This also matches the performance on individual kernels (e.g., Figure 5, 6, 7, 8).
This result indicates that APNN-TC can bring benefits for latency-sensitive applications.

\subsection{Additional Studies}
We perform several additional studies in this subsection, including the latency breakdown from individual NN layers and the benefit from kernel fusion.
We show results from RTX 3090 and skip results from A100 since we observe similar trend on these two GPUs.

\begin{figure}[t] \small
    \centering
    \includegraphics[width=0.46\textwidth]{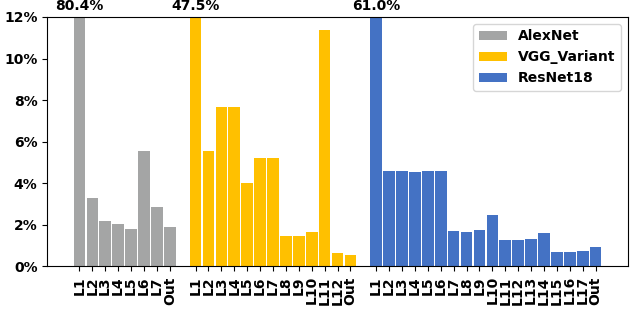}
    \vspace{-10pt}
    \caption{Per-layer latency breakdown of APNN models.}
    \vspace{-8pt}
    \label{fig: per-layer latency breakdown}
\end{figure}

\textbf{Latency Breakdown.} Figure~\ref{fig: per-layer latency breakdown} illustrates the percentage breakdown of the latency for the inference of 8 images over three NNs on RTX-3090 GPU.
Clearly, the first layer introduces the most delay since the input feature size for this layer is significantly larger than other layers.
This percentage can be as high as $80.4\%$ for AlexNet and $47.5\%$ for VGG\_Variant. On other layers, we observe a roughly balanced latency.
\begin{figure}[t] \small
    \centering
    \includegraphics[width=0.97\linewidth]{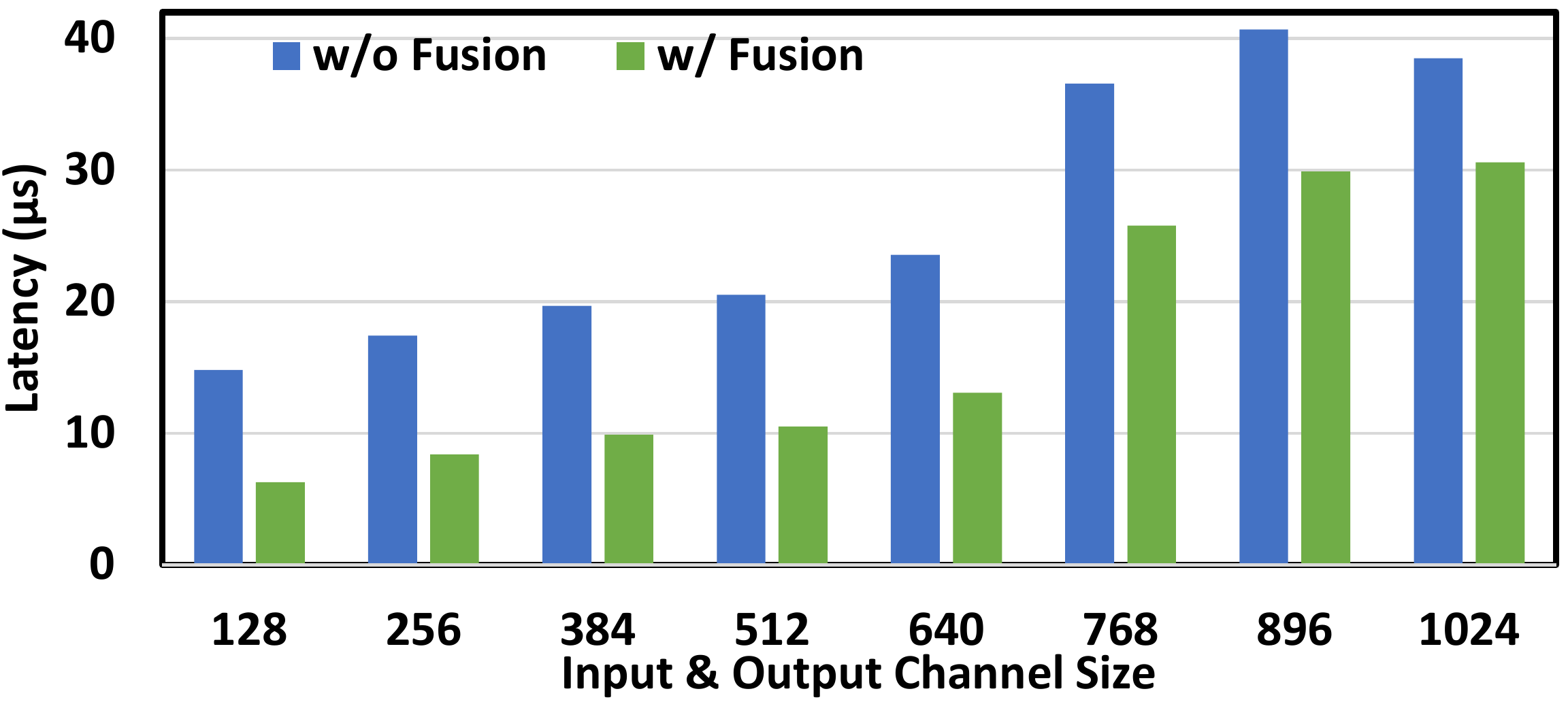}
    \vspace{-10pt}
    \caption{Speedup from APNN Kernel Fusion.}%
    \label{fig:KernelFusionBenefits}
    \vspace{-12pt}
\end{figure}

\textbf{Benefits from Kernel Fusion.}  
Figure \ref{fig:KernelFusionBenefits} investigates the performance benefits from fusing APConv-w1a2, pooling, and quantization into one kernel.
Specifically, in the "w/o Fusion" implementation, we implement three global functions for APConv-w1a2 with 32-bit output, $2\times 2$ pooling, and quantizing into $2$-bit outputs, respectively.
Here, each function read and write data to the global memory.
In the "w/ Fusion" implementation, we conduct the same workload in a single kernel.
Overall, we observe a latency reduction of $1.77\times$ on average.
The main reason is that, in "w/ Fusion", data across APConv, pooling, and quantization can be cached in shared memory and global memory access is significantly reduced.
\begin{figure}[t] \small
    \centering
    \includegraphics[width=0.97\linewidth]{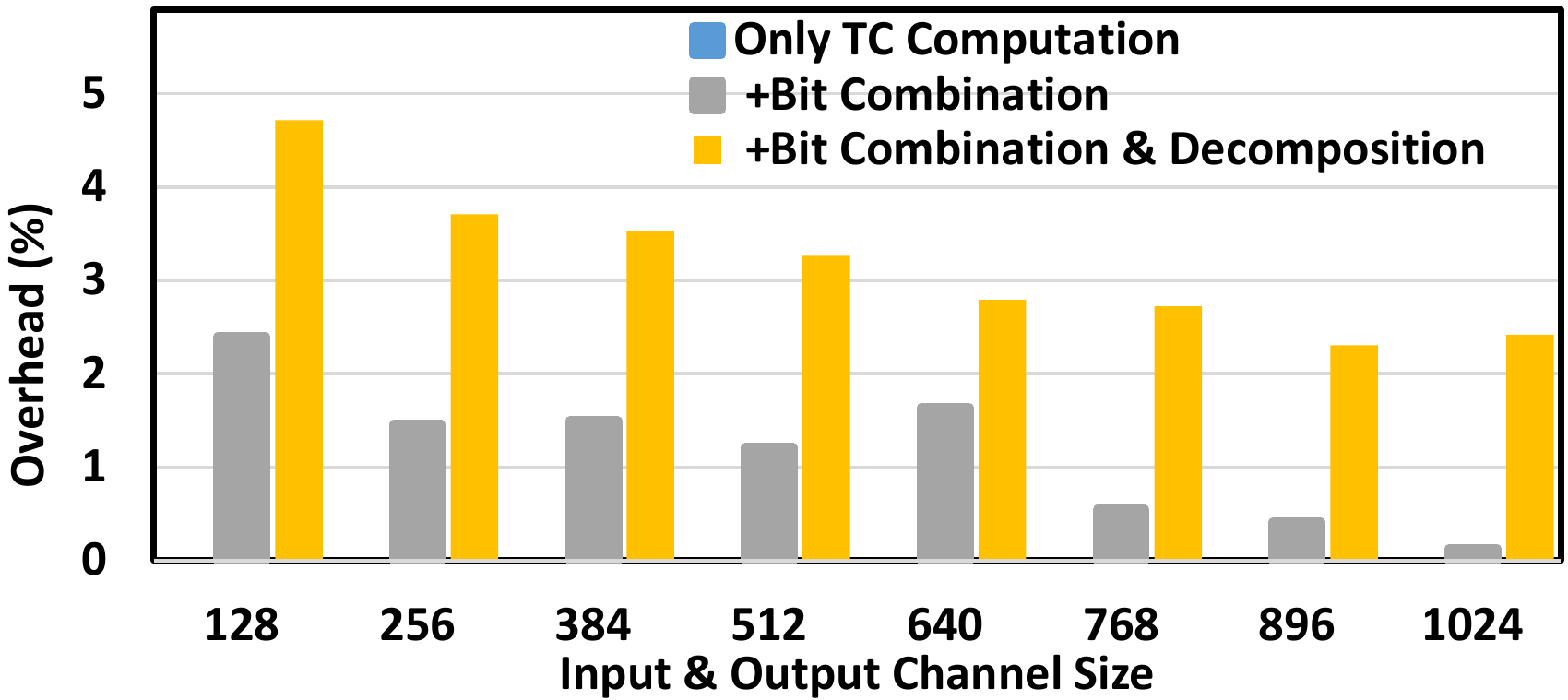}
    \vspace{-10pt}
    \caption{Overhead from bit combination and bit decomposition, relative to TC Computation.}
    \label{fig:bitOperationOverhead}
    \vspace{-10pt}
\end{figure}

\textbf{Overhead from bit combination and bit decomposition.}
We show the overhead from bit combination and bit decomposition in Figure \ref{fig:bitOperationOverhead}.
We profile the overhead on APConv designs following the same setting as Section \ref{sec:APConvPerformance}.
We show results from APConv-w1a2 since we observe similar overhead across bit settings.
On average, we empirically observe 1.16\% overhead from bit combination and another 2.02\% overhead from bit decomposition, compared to only TC computation.
The main reason is that bit combination and bit decomposition introduce only quadratic time complexity, which is significantly smaller than the cubic time complexity from TC computation.
Due to this difference in time complexity, the overhead from bit combination decreases from $2.4\%$ to $0.12\%$ as the channel size increases from $128$ to $1024$.
We also observe similar trend for bit decomposition.
\begin{figure}[t] \small
    \centering
    \includegraphics[width=0.97\linewidth]{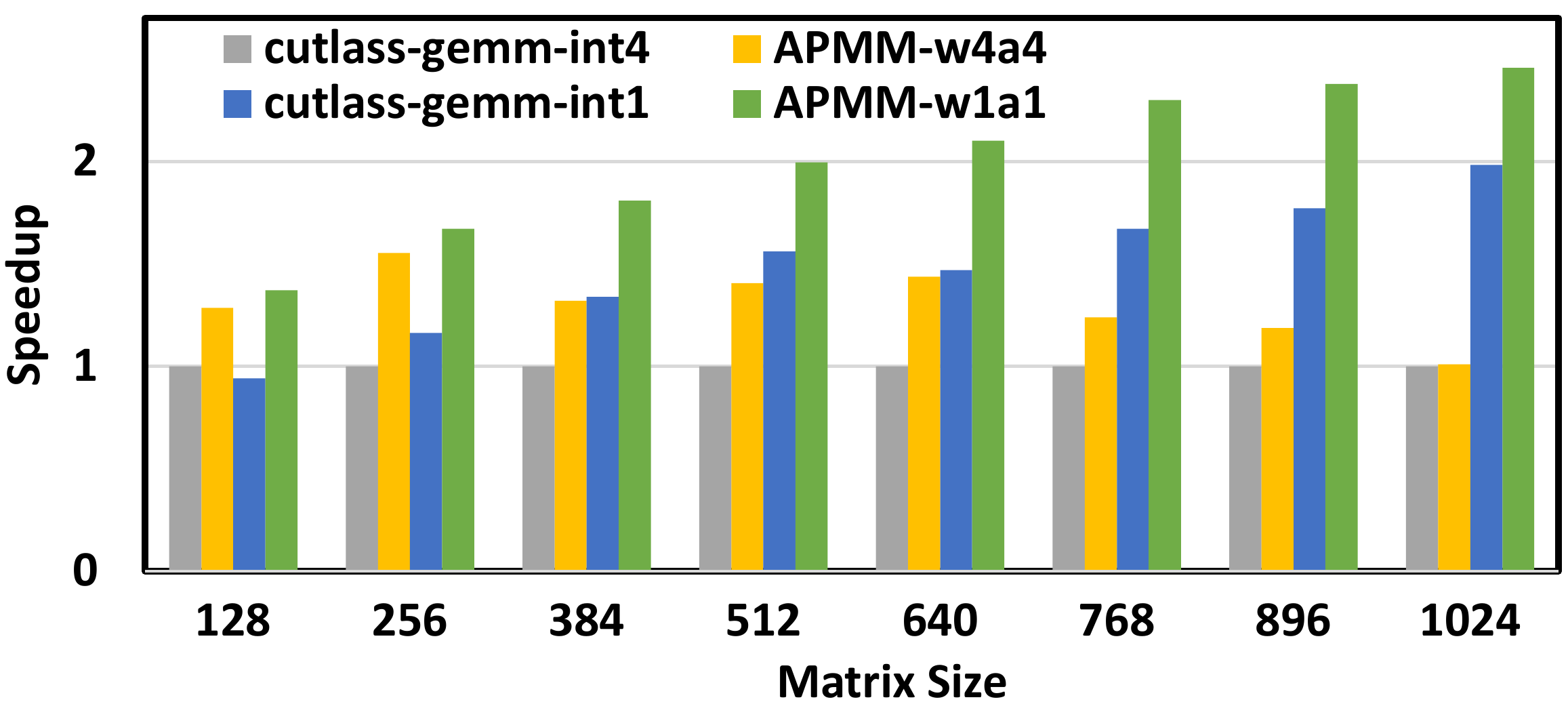}
    \vspace{-7pt}
    \caption{Comparing APMM and CUTLASS-GEMM.}
    \label{fig:sameBitComparison}
    \vspace{-7pt}
\end{figure}

\textbf{Comparing APMM and cutlass GEM under the same bits.}
Figure \ref{fig:sameBitComparison} shows the performance comparison between APMM and cutlass-gemm when using the same bits.
Overall, we observe that APMM-w4a4 can achieve $1.3\times$ speedup over cutlass-gemm-int4.
The main reason is that APMM-w4a4 can achieve better parallelism by using $16$ int1 computations to emulate $1$ int4 computation and achieving better GPU utilization, especially for small matrix sizes.
We note that this speedup of APMM-w4a4 over cutlass-gemm-int4 decreases as the matrix size increases where more int1 computation resources are required for emulation.
We also observe that APMM-w1a1 can achieve $1.35\times$ speedup over cutlass-gemm-int1.
This shows the benefit from our kernel-level optimizations.
\begin{table}[t] \small
    \centering
    \caption{Raw latency of a typical fully-connected layer with batch size $M=64$, input dimension $K=1024$, and output dimension $N=1024$. Unit: microsecond.}
    \label{tab:rawLatency}
    \vspace{-5pt}
    \scalebox{0.9}{
    \begin{tabular}{c|c|c|c|c|c}
    \hline
        \textbf{w1a2} & \textbf{w1a3} & \textbf{w1a4} & \textbf{w2a2} & \textbf{cutlass-gemm-int4} & \textbf{cutlass-gemm-int1} \\
    \hline
    \hline
        6.67 & 6.81 & 7.06 & 7.15 & 15.61 & 7.92 \\
    \hline
    \end{tabular}
    }
    \vspace{-5pt}
\end{table}

\textbf{Raw latency of a typical fully-connected layer.}
Table \ref{tab:rawLatency} shows the raw latency of a typical fully-connected layer with batch size $M=64$, input dimension $K=1024$, and output dimension $N=1024$.
Overall, we observe that we require only around $7$ microsecond for such a layer.
Comparing with cutlass-gemm-int4, we can achieve $2.27\times$ speedup on average by using arbitrary-precision computation.
We also note that the arbitrary-precision computation is even slightly faster than the cutlass-gemm-int1, which matches the result in Section \ref{sec:APMMPerformance}.

%% file: 07_Discussion.tex
\vspace{-3pt}
\section{Discussion}





\textbf{Practical usage of APNN.}
Arbitrary-precision neural networks have been widely studied to provide diverse tradeoffs between precision and efficiency \cite{HanMD15,TPDS,BinaryConnect,DoReFa,zhang2018lq,HAQ,OLCEL,li2019bstc}.
While arbitrary-precision may slightly reduce the precision, it shows merit in many practical usages such as smart sensors \cite{sensor,KungZWCM18,McDanelTK17}, mask detection \cite{BinaryCoP}, and intelligent agriculture \cite{GarofaloT0RB20}. In these usages, when a certain accuracy bar is surpassed, other essential metrics such as real-time processing and resource consumption are more important. For example, BinaryCoP \cite{BinaryCoP} utilizes low-power binary neural networks to detect facial-mask wear at entrances to corporate buildings and airports. Another example is XpulpNN \cite{GarofaloT0RB20} that uses quantized neural network on energy-efficient IoT devices.

\textbf{Generality to other NNs.}
This paper reports the results of APNN-TC on two most time-consuming kernels, GEMM and Convolution, from the computer vision domain and showcases the performance on popular vision models (e.g., AlexNet, VGG, and ResNet). 
Yet, we expect that APNN-TC applies to NNs from various domains such as natural language processing (NLP).
Intuitively, APNN-TC accelerates GEMM and dot products which is the building block of many NLP NNs \cite{Transformer,ZhangFB20,DevlinCLT19}, such as the attention layer and the feed-forward layer.




\textbf{Generality to other processors.}
APNN-TC utilizes population count (i.e., \texttt{popc()}) and two logical operations (i.e., \texttt{XOR} and \texttt{AND}) to support arbitrary-precision computation on Nvidia GPUs.
Considering the wide support for \texttt{popc()} and logical operations, APNN-TC can be easily adapted to diverse processors.
For example, AMD GPUs \cite{AMD} supports population count (i.e. \texttt{popcnt()} on AMD GPUs) and logical operations (e.g., bitwise \texttt{XOR}). Xeon phi \cite{Intel} also supports population count and logical operations.



%% file: 08_conclusion.tex
\vspace{-3pt}
\section{Conclusion}
In this paper, we design and implement APNN-TC that accelerates arbitrary-precision neural networks on Ampere GPU Tensor Cores.
Specifically, APNN-TC contains an int1-based emulation design on Tensor Cores to enable arbitrary-precision computation, an efficient AP-Layer design for efficiently mapping NN layers towards Tensor Cores, and an APNN design to minimize the memory access across NN layers.
Extensive evaluations on two Ampere GPUs show that \Mname~can achieve significant speedup over CUTLASS kernels and various mainstream NN models, such as ResNet and VGG.

%
%
%
%